\documentclass[12pt]{article}

\usepackage{amsmath,amsfonts,amssymb,latexsym,graphicx}

\numberwithin{equation}{section}
\def\be{\begin{equation}}
\def\ee{\end{equation}}
\def\bq{\begin{eqnarray}}
\def\eq{\end{eqnarray}}
\def\beq{\begin{eqnarray}}
\def\eeq{\end{eqnarray}}

\begin{document}
\title{100 years of mathematical cosmology:\\ Models, theories, and problems}
\author{\textsc{Spiros Cotsakis$^{1,2}$\thanks{skot@aegean.gr}, Alexander P. Yefremov$^1$\thanks{efremov-ap@rudn.ru}}\\
$^1$Institute of Gravitation and Cosmology, RUDN University\\
ul. Miklukho-Maklaya 6, Moscow 117198, Russia\\ 
$^2$\textsc{geo.dy.sy.c} Research Laboratory,
University of the Aegean\\ Karlovassi 83200, Samos, Greece}
\date{March 2022}
\maketitle
\newpage
\begin{abstract}
\noindent An elementary survey  of mathematical cosmology is presented. We  cover certain key ideas and developments in a  qualitative way, from the time of  the Einstein static universe in 1917 until today. We divide our presentation into four main parts, the first part containing important cosmologies  discovered until 1960. The second period (1960-80) contains discussions of geometric extensions of the standard cosmology, singularities, chaotic behaviour, and the initial input of particle physics ideas into cosmology. Our survey for the third period (1980-2000) continues with brief descriptions of the main ideas of inflation, the multiverse, quantum, Kaluza-Klein,  and string cosmologies, wormholes and baby universes, cosmological stability, and modified gravity. The last period which ends today includes various more advanced topics such as M-theoretic cosmology, braneworlds, the landscape,  topological issues, the measure problem, genericity, dynamical singularities, and dark energy. We emphasize certain threads that run throughout the whole period of development of theoretical cosmology and underline their importance in the overall structure of the field.  We end this outline with an inclusion of the abstracts of all  papers contributed to the Philosophical Transactions of the Royal Society A, Theme Issue `The Future of Mathematical Cosmology'.
\end{abstract}
\newpage
\tableofcontents
\newpage
\section{Introduction}
Before we begin,  we deem it fit to provide some comment on our choice of the somewhat restrictive word `mathematical' in the title of this paper. We shall use the phrase `mathematical cosmology' as almost indistinguishable to `theoretical cosmology',  ascribing to both an almost identical meaning. To us, the inclusion of material under both qualifications has an analogous interpretation and is closer in spirit to the two principles of selecting material for inclusion in their Course of Theoretical Physics used by Landau and Lifshitz (cf. \cite{ll3}, p. xi): We shall not deal with topics that may not be properly expounded without requiring  at the same time a detailed account of the existing observational results, and secondly, we shall not discuss too complicated applications of the theory. (Both of these criteria are of course somewhat subjective to some extent.)

We shall then use the phrases `mathematical cosmology' and `theoretical cosmology' as having a very similar meaning, a possible  difference being that of the amount of mathematical rigor contained in the presentation of the results. (We are well aware of the difficulties surrounding the general notion of `physical reality' as it arises in physics, some of which are probably aggravated in cosmology. Traditionally this notion is used to reflect differences between `theoretical' and `mathematical' physics, but we choose not to enter into such a discussion in this paper.) The emphasis in this  approach is perhaps  almost orthogonal to the more popular approach followed nowadays, the  data-driven, empirical,  or `actual  cosmology' (for the latter, see \cite{pee} for a standard treatment).

In the following Sections we shall present a panorama of cosmological models that aim at describing some aspect of the universe. But what is cosmological model? The  standard definition of a cosmological model in relativistic cosmology (see, e.g., \cite{emm} and refs therein), although extremely helpful for giving orientation and meaning to a very wide set of issues in cosmology, is perhaps too narrow for our present purposes.

Such definitions (or similar ones given in \cite{cotsakis, peter}) present  cosmological models as  having a timelike (fundamental) vector field together with a manifold and a spacetime metric which satisfies the Einstein equations. Apart from the fact that general relativity is currently only one out of many different geometric ways of describing an interaction of spacetime geometry and matter, there are areas in cosmology where even the very notion of spacetime and that of a fundamental timelike field are probably too restrictive to use. As typical examples of this, we may mention the multiverse,  the landscape, or the notion of braneworld, all  of which require other types of geometry and dynamics for their description rather than that found in the standard definition of cosmological model.

The beauty and enormous variety of ideas of modern mathematical cosmology  have their roots in the different kinds of geometry required to be developed and studied in parallel and in conjunction with those cosmological  ideas  for a better understanding of  different aspects of this most majestic of fields of theoretical physics.

The purpose of this paper is twofold. Firstly, we describe some important key developments in the field of theoretical cosmology since its modern beginning in 1917. We  single out and discuss some of the important ideas that characterize the nature of this field within theoretical physics. Secondly, we provide a short description of the contents of the Theme Issue `The Future of Mathematical Cosmology', by providing the titles and abstracts of the  particular contributions.

The plan of this paper is as follows. In the following 36  subsections contained in Sections 2-5, we give brief descriptions of the emergence, development, importance, and interconnections, of most major subfields  of theoretical  mathematical  cosmology, such as relativistic, modified gravity and dark energy,  inflationary, quantum,  string, M-theoretic, and brane cosmologies; we also discuss  alternative domains and threads of development such as singularities, horizons, measures, stability, genericity, and topology.    We focus on  key theoretical discoveries  as well as  fundamental ideas that were to become (or,  in fact, may become) instrumental in the development of the whole field. This is presented in four key time periods ranging from 1917 to today.  In this perspective, at the end of this paper, we also comment on  the contents of individual invited contributions to the Theme Issue `\emph{The Future of Mathematical Cosmology}' that are included in two separate volumes of the Philosophical Transactions of the Royal Society A.

\section{First period, 1917-1960}
In the first period of developments of mathematical cosmology, there appeared a great number of novel fundamental ideas and still play a major role today in shaping the field of theoretical cosmology:
\begin{itemize}
\item The `universes' as  solutions of  the field equations
\item The cosmological constant and vacuum energy
\item Homogeneity of the universe
\item Inhomogeneous and anisotropic cosmologies
\item Expansion and contraction
\item Evolution vs. steady state
\item Big bang vs. bouncing models
\item Gravitational stability and perturbations
\item Hot big bang
\item Causality and time travel
\item Local vs. global structure
\end{itemize}
All these ideas are as fundamental and important in theoretical cosmology research  today as they were in the beginnings of modern cosmology. It is an amazing fact that the subsequent development of the field in the coming decades proved that all these novel and important ideas are still playing a major role, but at the same time are a small part of the complex network of methods and directions that constitute mathematical cosmology today.

\subsection{Einstein static universe}\label{static}
In 8 February 1917, Einstein announced the first application of general relativity to the universe considered as a whole, the \emph{Einstein static universe}\footnote{For reprints of  early papers appearing in the text for the next few subsections, see \cite{bern}. For an early history of  theoretical cosmology in the 20th century, see \cite{north}.} \cite{ein0}. It was a cosmological model of motionless matter, having  positively curved spatial sections (so had a finite volume but no boundary), and an added repulsive force (dependent on the cosmological constant) balancing the attractive gravitational force, resulting in a static world. But perhaps the most distinctive new feature of Einstein's static universe was its assumed overall homogeneity. In other words, in Einstein's proposal, there is an upper bound - provided by large-scale homogeneity - to the hierarchical structure in all of physics starting from elementary particles, atoms, molecules, planets, etc, and ending with large-scale uniformity. (For the opposite view, that of \emph{no} upper bound to the hierarchy of clusters of galaxies, see Section \ref{fractal}.)

Einstein's static world remains always static because of the presence of Einstein's cosmological constant - the $\lambda$ term - a new, additional term in the original  Einstein's equations of general relativity that he introduced for that purpose. The spherical nature of Einstein's static universe implies a maximum (spherical) distance around the universe, the so-called circumnavigation time, equal to $t_{ct}=2\pi a$, $a$ being the scale factor (or, radius of curvature). Then from the field equations, it follows that $t_{ct}\sim\sqrt{4/\rho}$ hours (in suitable units),  so that this time is increasing with decreasing density, being equal to about 2 hours for the water density. The propagation of light in an Einstein static universe (as seen in the usual spacetime diagram depiction of it, the `Einstein cylinder'), implies that this model acts like a lens, antipodal points are seen nearby to each other by observers situated in them, and receding images from a point in this universe first appear smaller until they reach its antipode, when they look bigger.

But perhaps the most important property of Einstein static universe is its instability against homogeneous and isotropic perturbations discovered as well as  other types of fluctuations a few years later (see below). This important instability property reveals that Einstein's world can only be one of a \emph{transient nature}, and depending on its matter content, it can connect with other eras of the cosmological history (see Section 4).

Einstein's idea was truly extraordinary. By extension, every solution of the field equations of general relativity would describe some possible universe, and here was the first one. This discovery marked the beginning of cosmology: the `universe' - in this case Einstein's static solution - comes equipped with a variety definite `physical' properties as predicted by general relativity, it is homogeneous (and isotropic), finite, spatial closed (with topology that of the 3-sphere),  positively curved, having a nonzero matter density, and static.

This led  Einstein to struggle with the dilemma: There was apparently only one Universe, but an infinite number of \emph{universes}, that is unknown  \emph{solutions} to the mathematical equations of general relativity describing possible models of the Universe; so how could one reconcile the two? Also how was it possible to control the universes' behaviour at infinity, if that was allowed elsewhere, or avoid having a finite edge? A vast field of investigation had just opened.

\subsection{De Sitter spacetime}
Very soon the second universe, solution of the modified Einstein equations with the lambda term,  was found by W. de Sitter \cite{deS0}. That described an empty (zero matter density) space, static model as was originally proposed by de Sitter.  It   was later realized that it may also be described a model for an expanding universe in both time directions,  because of the negligible effect of gravitational attraction. In fact, it is constantly accelerating with time due to the Einstein's cosmological constant term introduced earlier the same year that de Sitter spacetime first appeared.

The \emph{de Sitter universe} has no beginning or end. This universe plays the role of an `undisturbed' state for the modified Einstein equations with a cosmological term, in a similar way as does Minkowski space for the original field equations (i.e., with $\lambda =0$). However, de Sitter space is really different than Minkowski space in certain quantum aspects, see below. It also plays a fundamental role as possible asymptotic states of more general homogeneous cosmologies (see later Sections below).

In fact, there are two `de Sitter' spaces, one for $\lambda$ positive - \emph{the} de Sitter space (dS) - and the case of negative  $\lambda$, the so-called anti-de Sitter space (AdS) (see \cite{rin1} for an excellent discussion of these spaces, and \cite{he} for  a more specialized description of their infinity properties).

Perhaps the simplest way to distinguish dS from AdS universes is  that dS, like Schwarszchild spacetime but unlike AdS, has an horizon separating its static regions from its evolutionary ones. This was not clear initially, but it later became understood with the work of Eddington and Lema\^{i}tre. However, unlike dS space,  the AdS spacetime contains closed timelike curves and is globally static. Normally, when we talk about AdS we simply mean its universal cover, a space without closed timelike curves.

dS and AdS spaces are very important in current primordial cosmology, especially in braneworlds and holographic models of the early universe (see Section \ref{brane} below). dS space also has an instability, classically forbidden, associated with the possibility of defining a non-thermal temperature using the event horizon \cite{gh77}. In this case one may use the Schwarzschild-de Sitter (Kottler) solution that describes a black hole immersed in de Sitter space, to construct Euclidean instanton solutions with special properties (cf. \cite{pe} and refs. therein).

\subsection{Fractal universes}\label{fractal}
Einstein's static universe was the first application of general relativity to cosmology and, in fact, it created such a boost to the study of universes that an application of the old Newton's theory to cosmology seemed an impossibility. However, in 1922 the claim of Einstein of an upper bound to the galaxy hierarchy was to be challenged  by C. Charlier \cite{cha}, who introduced a \emph{fractal universe} as it may now be called - a distribution of a never-ending repetition of clusters \emph{ad infinitum}, on larger and larger scales (the word fractal was introduced much later \cite{ma1}).

Charlier's universe possessing a fractal hierarchy was based on earlier ideas of F. d'Albe and others (see \cite{har} for an early history of such ideas), and was arranged such that,  although the overall density and volume were infinite, the \emph{average} matter density was zero, thus bypassing the obstacles of the `gravitational paradox'. (According to the Newton-Ivory theorem  \cite{arny},  at infinite distance in an infinite universe of constant density, the Newtonian force may take arbitrary values.)

Also, it has the same clustering on all scales, a property avoiding the difficulties associated with Olbers' paradox, that is the integrated luminosity of all stars at every place in the universe is now finite - the night sky is still dark. In addition, the gravity pull is everywhere finite and so star velocities remain small at all scales.

Charlier model of an inhomogeneous, infinitely large, fractal universe with infinite mass but zero average density, was soon after treated in a more elaborate way by F. Selety \cite{se1}, and more recently by others \cite{devac,dy}, but an objection (originally raised by Einstein) was how such a hierarchy could arise in the first place, remained unresolved.

More recently, attempts to explain the large-scale clustering  by a fractal universe have re-emerged \cite{labini1,labini2,labini3}, although conflict of such explanations with the  large-scale observed homogeneity from the temperature fluctuations in the cosmic microwave background (hereafter CMB) is  taken as an indication that clustering cannot continue indefinitely.

\subsection{Friedmann solutions}
The  true gallery of possibilities allowed by the general theory of relativity was first discovered by A. Friedmann in 1922-24 \cite{fri1,fri2}. Friedmann discovered  a family of universes that were  homogeneous and isotropic, having matter,  non-zero curvature, as well as nonzero cosmological constant. He found out that they are described  by equations containing the previously found solutions as well as many other novel additional universes. (Today Friedmann universes are also called Friedman-Robertson-Walker, Robertson-Walker, Friedmann-Lema\^{i}tre, or even Friedman-Robertson-Walker-Lema\^{i}tre solutions.We shall use the notation `FRWL' below.)

The non-static character  of the \emph{Friedmann universes} fitted well with 1912 observations of V. Slipher \cite{sli} showing a red shift  in the wavelengths of distant nebulae (later coined galaxies), as did de Sitter's universe, but  no one at the time then was ready to accept these results as showing an overall expansion of the universe. These universes are models with source a perfect fluid, and may or may not have a cosmological constant (the name Friedmann-Lema\^{i}tre is sometimes  reserved for the latter case).

\subsection{Lema\^{i}tre universe}
The connection of the dynamical character of Friedmann models with the observations showing an overall expansion was done a few years later. G. Lema\^{i}tre in 1927 \cite{le1} combined the Friedmann models with the phenomenon of redshift, calculating for the first time the constant in the `Hubble's law' $v=Hr$, connecting the distance to recession velocity, found by Hubble two years later.

The 1927 paper of Lema\^{i}tre was translated into English in \cite{le1a}  after a suggestion of Eddington   (for this amazing story that made Lema\^{i}tre `the most well-known theoretical cosmology of the day',  see \cite{bookofuni}).  In that same 1927 paper, Lema\^{i}tre had also shown that the Einstein static universe was unstable, small irregularities would in fact grow in time. The resulting  \emph{Lema\^{i}tre universe} has the property of coasting as static an infinite time in the past, before becoming expanding and accelerating like de Sitter's universe in the future (this instability of the Einstein static world was also shown independently a few years later By A. Eddington \cite{edd1}, and this universe is sometimes called the Eddington-Lema\^{i}tre cosmology).

Sometimes by Lema\^{i}tre universe we mean one that starts with a big bang (reserving the infinite coasting static period for the  Eddington-Lema\^{i}tre model), then passes through a finite hesitation period where it is described by the Einstein static universe, and then ends with  a `whimper'   expanding like a de Sitter space into the future.

Lema\^{i}tre also appears the first to connect the cosmological constant to the energy of the vacuum in a paper published in 1934 \cite{le2}, now related to what is termed dark energy.

\subsection{Einstein-de Sitter universe}
But the situation with a zero cosmological constant was no less interesting. Einstein and de Sitter jointly on 1932 emphasized the importance of what is perhaps the simplest one of all the Friedmann solutions, the \emph{Einstein-de Sitter universe} as it is now called \cite{ein-de}. It is a flat model starting a finite time in the past, and expanding at a critical rate with scale factor proportional to $t^{2/3}$ and having zero cosmological constant.

Today this model agrees with observations so greatly that it has become a problem to explain why this is so and why instabilities have not ruined it one way or another (`flatness problem').

\subsection{Tolman's bouncing universe}\label{tolman}
Another distinctive aspect of some of the Friedmann solutions with zero lambda is their symmetric or \emph{cyclic character}, as they evolve from small to large to small, and a natural question suggested itself the possibility of an infinite number of such cycles.

This was taken seriously in \emph{Tolman's bouncing universe} of 1932 \cite{tolman}, who discovered that such a possibility, in the case of zero cosmological constant,  must be accompanied by an increase in the overall entropy of the universe from cycle to cycle, increasing the maximum of the expansion in a cycle relative to the previous one. This is in sharp contrast to the idea of a beginning in a finite time in the past and end in the future.

Oscillating universes like Tolman's play an important role in theoretical cosmology today. However, the main issues that should be addressed in such models is the ever-increasing temperature of the CMB from cycle to cycle, as well as the amount of radiation and the entropy. We shall see later the Tolman's model may be useful to motivate the existence of wormholes.

This concludes the various properties of the Friedmann universes which, although being the simplest class of cosmological model, is one of the most important and must be thoroughly understood in any cosmological theory before embarking on the study of more complicated models.

\subsection{Kasner models}
A novel feature of possible universes with no cosmological constant was discovered by E. Kasner as early as 1921 \cite{kasner}, an expanding universe starting a finite time ago but having the property of \emph{evolving in time at different rates in the three different spatial directions}. Kasner's empty and flat universe may contract in one direction while expanding in the other two, keep a constant volume. It is \emph{anisotropic}. In the past direction, neither matter not a lambda can alter its behaviour and avoid a singularity, but in the future a matter-filled Kasner model looks like Einstein-de Sitter, while a `lambda-Kasner' universe looks like de Sitter.

In addition, the Kasner universe plays a significant role in many parts of cosmology. Its late-time asymptotics (as time approaches infinity) are isotropic, as they approach the Einstein-de Sitter, and so Kasner may be a suitable model for the real universe. At early times however, it contracts to the singularity anisotropically: it can do so either with two of the orthogonal directions contracting and one expanding, resulting in a \emph{cigar} singularity. On the other hand, if two directions are expanding and one contracting, the model develops a \emph{pancake} singularity. These kinds of singularity  are absent in the simpler FRWL models. We also note that in this special case, the expanding direction has the property of allowing communication around it in a cylinder, the model develops no horizons along that direction.

The Kasner universe is the simplest cosmology of  a  much wider family of cosmological models, the Bianchi universes, which we  discuss below. It is very important in current studies of the cosmological singularity because it appears that some form of `Kasner behaviour' must be taken into account in describing the singularity.

\subsection{Lema\^{i}tre-Tolman universe}
Then a joint development  really took things at the next level. A common feature of all universes discussed so far (except the fractal hierarchical universe) is their large-scale homogeneity, either in the isotropic FRWL universes, or in the anisotropic Kasner spacetime. But the first discoveries of \emph{inhomogeneous  universes} soon came along.  The \emph{Lema\^{i}tre-Tolman universe} \cite{l1,t1} contained under- and over-densed regions than some average density, with  the rate of expansion, curvature and matter density depending on the spatial points, making the overall evolution of the universe more and more inhomogeneous with time.

These universes would locally collapse to form galaxies (or `...dust particles (nebulae) which exert negligible pressure...' as Tolman calls them in \cite{t1}), and being globally inhomogeneous would look totally different than our local homogeneous patch.

A similar inhomogeneous universe was found by Einstein and E. Strauss in 1945 \cite{es1}, the so-called \emph{Swiss-cheese universe}, a spherical, pressureless model in which spherical regions were removed and equal  masses were put back at the centers of the removed regions. Because of spherical symmetry there are no gravitational waves in this model, but this feature was later rescued by replacing with cylindrical symmetry, as in the Einstein-Rosen later solution \cite{er}.

More generally, the importance of inhomogeneous solutions in cosmology cannot be overestimated. One sees that even the earliest known solutions like the Swiss-cheese model introduce a level of sophistication higher than the simplest homogeneous solutions, in that they introduce new problems not met before, but also the problem of how to join the solutions at their common boundary, \emph{ the junction conditions.} This  in turn plays a key role in many areas of modern cosmology, such as the deformation geometry and evolution of strings, branes, and higher-dimensional  membranes propagating in spacetime (cf. e.g., \cite{cg1}, \cite{ca1}, and refs, therein), not to mention  quantum cosmological aspects (cf. e.g., \cite{mann}).

\subsection{Lifshitz universes}
The exact solutions considered in the previous subsections were useful in turn because they pointed to the general problem of how to combine anisotropies as in the Kasner world, and inhomogeneities as those present in the models of Einstein-Rosen-Strauss.

Mathematically this is related to the problem of how to carefully study  all possible  \emph{small perturbations} of the homogeneous and isotropic universes of Friedmann. This was first done in \emph{E. Lifshitz's theory} \cite{lif} of first order cosmological perturbations, a landmark study in 1946. Lifshitz's analysis revealed that the Friedmann solutions have the hidden property of \emph{gravitational instability}; they share atypical, nongeneric features, not present in a more  general situation. A further analysis that appeared later treated not only fluctuations inside the horizon (that is smaller than the radius of curvature) but also those comparable to it, cf. \cite{bkl1}, \cite{ll2}, Sect. 115). For the second-order theory and recent work, see \cite{tom,rio}, and for an advanced textbook treatment \cite{weinberg2}, chap. 5.

This study also set the stage for all future research on the problem of the origin and development of small cosmological fluctuations, arguably perhaps the most central problem of contemporary cosmology. This is so because through the field equations it is the basis of the issue of the formation of structure in the universe, namely, how irregularities that  originate in the early universe will develop coherently in the universe starting from small initial disturbances. This is a very involved issue connected with other major and not completely settled problems such as the horizon and flatness problems (for a thoughtful presentation, see, as for many other topics in cosmology, the excellent `introductory' treatment in the book \cite{har}).

\subsection{The hot big bang}
But at the time of the inhomogeneous studies of Lifshitz,  even in the homogeneous picture the last word had not been said yet. It was not until 1948 that G. Gamow \cite{ga1} and two of his students, R. Alpher and R. Herman,  proposed that \emph{the early universe was not only very dense but also very hot} and showed that in such a universe the ratio of the matter density to the cube of the temperature of  heat radiation  is a constant (see \cite{ga2,ah1} for an insightful presentation of their findings).

Hence, if one measured that density and ratio today, they could find the present temperature of the leftover radiation. The authors gave an estimate of this temperature about  5 degrees Kelvin, but no one gave any serious attention to this result back then, and went completely unnoticed,  despite that this prediction was no other than that of the later discovered cosmic microwave background thermal radiation left from a very early stage, the \emph{hot big bang}, in the evolution of the homogeneous universe.

This is a tremendous result equivalent to the discovery of a radiation era in the early  universe with density exceeding that of matter. Also this radiation survives until today with a much cooler temperature.
It took more than 15 years until the cosmic microwave background radiation, observationally discovered by Penzias and Wilson in 1965, convincingly explained by R. Dicke and his group the same year, and accurately measured to be isotropic as late as 1967.

This discovery was perhaps the single most important one in theoretical cosmology as it established it as  a physical science (see \cite{pee}, chap. 4,  for a very detailed analysis of the various developments in the long period leading to  this discovery). It also led, with the help of particle physics,  to \emph{the standard model of cosmology}, where a detailed history of events during the various periods in the history of the expansing universe may be built. This history extends from the period of the early universe (sometimes this period is also called 'the big bang') for  about 55 orders of magnitudes from the Planck time until the beginning of the current matter dominated era which lasted only for about 5.5 orders of magnitude (steps measured in logarithmic time). For a popular account of the standard model of cosmology, we refer to \cite{wein3}, and to \cite{peter} for a detailed presentation.

\subsection{Steady-state universes}
However, not all ideas during the first period in the development of theoretical cosmology were about the big bang. The steady state universe of \emph{continuous creation} by H. Bondi, F. Hoyle, and T. Gold in 1948 \cite{bg,hoyle} introduced the so-called perfect cosmological principle, a variant of the homogeneity principle that Einstein had introduced earlier in his static model, in which the universe looks the same not only in space but also for all times.

That was a completely different model from any big bang theory, eventually based on a different geometric principle we shall later analyze, the scalar-tensor theory (see \cite{nar1} and refs. therein, for a discussion of  steady-state models from that angle).

Not being static but `steady' (or stationary), that is eternally expanding with constant acceleration, the state of the universe required  matter to be continuously created in time, instead of it being created in a unique time as in the big bang model (incidentally the name `big bang' was coined by Hoyle during the 1949 in a BBC radio program). Such `creation theories' have matter to be created not from radiation but apparently out of nothing. This is to be compared with modern attempts  to describe spontaneous creation `out of nothing'.

Steady-state cosmologies sharing the feature of continuous creation include the one by W. McCrea proposed in 1951 \cite{mac1}, see also \cite{mac2}, that incorporated this idea within general relativity as cosmic tension (negative pressure, $p=-\rho$) leading to a global state of constant density. This ingenious argument of McCrea implied that expansion continuously  creates new matter to keep a constant mass density. (This is again to be compared with the inflationary idea of expansion at constant density coming from a  false-vacuum.)

These ideas were considered as rivals to the Friedmann models for a period, until it was found in 1952 by W. Baade that previous astronomical calculations of distance calibration were in fact wrong by a factor of two or more, a fact that erased the initial motivation for such alternative models. For a modern reincarnation of the steady-state idea, see \cite{pen1}.

\subsection{G\"{o}del's  universe}
Another important advance in theoretical cosmology before the 1960's was \emph{K. G\"{o}del's rotating universe} published in 1949 \cite{goe}, a zero-pressure fluid solution to the Einstein's equations with a negative cosmological constant. This solution for the first time allowed time travel into the past in exchange with relativistic speeds and exotic matter configurations.

A failure of a cosmic time function existing  for all fundamental observers (able to synchronize their clocks) implies that $t=\textrm{const.}$ hypersurfaces in spacetime cannot be spacelike everywhere, and will therefore turn timelike somewhere. This leads to causality violations, such as closed timelike curves crossing such hypersurfaces an odd number of times,  and in G\"{o}del's model this situation is generally provided by a non-vanishing vorticity in the timelike direction, cf. \cite{tip,he,ear} for more details.

More recently, however, some authors have extended the original solution to \emph{causal  G\"{o}del universes}. In these cosmologies closed timelike curves are absent in G\"{o}del-type models in geometric extensions of general relativity such as when a massless scalar field is added \cite{re}, in a higher-order gravity theory quadratic to the Ricci curvature \cite{ac}, and in various string theories \cite{bada}, or even hidden behind a holographic screen \cite{hor}.

G\"{o}del's universe was instrumental in the further development of modern theoretical cosmology because, despite its incompatibility with the observations about the expansion of the universe, clearly provided the first example of the basic distinction between local and global properties in cosmology.

\subsection{The many universes}
An efficient way to classify cosmological models is via the consideration of an isometry group acting on spacetime, thus defining an orbit through each spacetime point, i.e.,  submanifolds in spacetime where all physical and geometric properties remain invariant. Isometry, or symmetry, groups signify how symmetric a given cosmology is.

Roughly speaking, the dimension $r$, of an symmetry group acting on a  spacetime of dimension $n$,  splits into a homogeneous part - the orbit of dimension $d$ - and an isotropic part of dimension $s$, such that $r=d+s$. In this situation, $d$ signifies the degree of inhomogeneity, and $s$ that of isotropy.

For example, in a 4-dimensional spacetime ($n=4$), $d=4$ implies an unchanging world in space \emph{and} time, $d=3$ a spatially homogeneous universe, and for $d\leq 2$, that is $d=2, 1, 0$,  we have inhomogeneous universes, the last case corresponding to the real universe - no symmetry. For $s=3$, we have complete isotropy, $s=1$ is the so-called locally rotationally symmetric (LRS) universes, while $s=0$ is the case of no isotropy (the case $s=2$ is not possible). For a textbook treatment of the classification of universes in terms of a symmetry group (see for example, \cite{emm}, Part 4, and refs. therein).

A. Taub \cite{taub} studied the Einstein equations using group theoretic ideas for the first time in 1951. He exploited a classification of spacetimes into nine families  discovered by L. Bianchi in 1897, the so-called \emph{Bianchi universes} \cite{bianchi} (this is the case $s=0, d=3$). He discovered that all previously known homogeneous FRWL  solutions of the field equations could be easily accommodated in that classification, but there were many more solutions, like those discussed above, whose cosmological study were to occupy the coming decades.

One new feature of the Bianchi universes was the possibility of gravitational waves, as well as new types of early- and late-time evolution. These universes are all homogeneous but anisotropic, the Kasner model is described by the simplest one of them, the Bianchi type I, and so could accommodate other features not present in simple isotropic ones, like shear, vorticity, and anisotropic curvature, with respect to an overall isotropic Hubble flow.

The Bianchi cosmologies play also a role as asymptotic states of more general and more realistic inhomogeneous spacetimes. One particularly interesting class of exact inhomogeneous cosmologies are the Szekeres-Szafron models which may be regarded as nonlinear perturbations of the FRWL universes.

The real difficulty in dealing with inhomogeneous models is their different geometry. For an introduction to inhomogeneous cosmology see \cite{emm}, chap. 19, and for a full discussion of the known inhomogeneous models see \cite{kra}. In these situations, one also needs to have a well-defined, useful and general formalism for cosmological models. This issue has long been an important one in mathematical cosmology, as it is not a good practise to invent a new formalism every time one studies a different cosmological model. For an introductory discussion of the most well-developed formalism for general cosmology,  see \cite{emm}, Part 2, and also \cite{uvwe} and refs. therein.

\section{Second period, 1960-1980}
Novel ideas that emerged in this period with special relation to cosmology can be collectively summarized as follows:
\begin{itemize}
\item Geometric extensions of  general relativity
\item Singularity theorems, global techniques
\item BKL conjecture
\item Mixmaster universe and horizons
\item Particle physics and  classical singularities
\end{itemize}

\subsection{Brans-Dicke theory}
Building on earlier and more speculative ideas of P. Dirac on the variation of Newton's gravitational constant $G$ \cite{di1}, Brans and R. Dicke \cite{bd1} in 1961 came up with an ingenious theory of gravitation and a family of universes that proved cosmology was possible outside the realm of general relativity. The \emph{Brans-Dicke gravity theory} predicted universes which were expanding and contracting pretty much as in Einstein's theory, but their properties depended on the values of gravitational constant $G$ and its time-dependence through a new scalar field in addition to the metric, the primary function of which is the determination of the local value of $G$.

This scalar field is of purely cosmological origin, as it is not one of the curvature invariants formed by the curvature tensor which fall off more rapidly than $r^{-1}$ from a mass source, and so become unsuitable for cosmology as they are determined by nearly rather than distant matter.

This is  a distinctive feature of this theory, the mediation of  gravitation at cosmological distances by a scalar `degree of freedom' in addition to the tensorial degrees of freedom, resulting in three degrees of freedom rather than two  in general relativity. So one does not expect to find this scalar field in solar system tests, because it is supposed to be of a cosmological origin (in the same way that one does not apply solar system metrics to cosmology), cf. \cite{di2}. However, of course, solar system tests constraint the values of the BD parameter $\omega$ to values bounded below by 6 (cf. \cite{di1} Eq. (36)), or higher in more recent studies. It is also important that in BD theory $\omega$ be positive and of order one if contributions of nearby matter to the inertial reaction are to be positive \cite{di1}.

As a rule, any solution of general relativity with matter having energy-momentum tensor with a vanishing trace is a solution of the  Brans-Dicke theory. In addition, flat perfect fluid solutions approach the vacuum solutions at early times, whereas the late-time behaviour is `Machian' in the sense that matter dominates over the kinetic energy of the scalar field \cite{nari1}.

For anisotropic extensions of BD cosmology including the possible singularity removal, see \cite{rf1}, for an early description of the possible (non-)occurrence of oscillatory behaviour in BD models where a vector field in addition to the BD scalar is included see \cite{bkl4,r76}, and for more recent qualitative approach to dynamical stability in the space of all (isotropic) solutions, see \cite{ke1}. For interesting reviews of the early history of BD theory with much wider commentary and ideas on the influence and importance of scalar fields in cosmology, see \cite{brans1,brans2}.

This was the first example of a cosmological theory  qualitatively distinct from general relativistic cosmology, still perfectly reliable. The Brans-Dicke gravitational field is \emph{mediated} by a scalar field in addition to the spacetime metric, and Brans-Dicke theory has the property of being  equivalent to general relativity under a conformal transformation of the fields \cite{di3}. In this `Einstein frame', in distinction to the original frame of the Brans-Dicke equations - the so-called Jordan frame' - the theory appears in the conventional form with the scalar field playing the role of a matter field. Then the rest masses of all particles are affected by their interaction with the scalar field, which reduces thus their masses. This led to a novel interpretation of the smallness of  $G$: this  is interpreted as  small because the masses are conformally  reduced drastically  by their interaction with the field which is generated by  all the matter at cosmological distances \cite{di3}.

In any of these two interpretations, the scalar field gives a time-dependent Newton's constant $G$, and so Brans-Dicke theory appears as a framework for the construction of concrete models for the time-variation of the `constant' $G$ during the evolution of the universe. For a discussion of the new and intriguing possibilities of varying constants in theoretical  cosmology, see \cite{ba1,ba2}.

Starting in  the 1980's, and perhaps with the input of string cosmology, other cosmological extensions of general relativity began to attract serious attention. The whole family of possible gravitational actions  prevailed under the name \emph{scalar-tensor cosmology}. Based on early work of P. G. Begman, R. V. Wagoner, and K Nordtvedt on extensions of the Brans-Dicke prototype theory, various possibilities appear with their myriads of possible couplings between matter fields, scalar fields,  and gravity (see, e.g., \cite{dn1,mn1} for the `relaxation' problem of such theories to general relativity, with refs. to earlier work).

A basic characteristic of scalar-tensor theories is that the original BD dimensionless coupling constant  $\omega$  is promoted to a dimensionless coupling \emph{function} of the scalar field, thus hugely enriching the dynamical possibilities (and problems!) of the original BD theory. Scalar-tensor theory, in particular `pure' BD theory  also appears to be  related to string theory, the latter predicting however, a non-minimal and non-universal coupling to the BD scalar (coined the `dilaton' in string contexts), cf. \cite{gasp}.

\subsection{Torsion and cosmology}
A different type of geometric alternative to general relativity with additional fields that appeared very early in the 1960's  was the \emph{Einstein-Cartan gravity theory}, a mixture of the structures of general relativity with another theory introduced by Elie Cartan \cite{cartan} in 1924, and based on two invariants of the Poincare group - the space-time curvature and \emph{torsion}. In solid-state physics torsion was used to describe dislocations in crystals, however, attempts to include torsion in Einstein's already well-studied and developed theory of gravity appeared only in the 1960's. The first people who recalled torsion were, apparently, Kibble \cite{kibble} and Sciama \cite{sciama},  in whose works an original idea was voiced about the possible relationship of space-time torsion with the proper angular momentum of matter.

But the theory began to develop only after the first (and sensational) torsion effect was calculated. In the works of researchers of the Krakow school Kopczynski \cite{Kopczynski} and Trautman \cite{Trautman}, it was demonstrated that the torsion of space-time, which geometrically reflects the fact of polarization of spin of the dust particles, the source of the gravitational field, can eliminate cosmological singularities in  Friedmann universes. The Einstein-Cartan theory has two essential features: (i) the torsion field equations (obtained by varying the Lagrangian by affine connection) are purely algebraic equalities linking torsion with its source, the spin of particles of gravitating matter; this means that torsion does not extend in this theory. (ii) Since the same Lagrangian (scalar curvature) with a single coupling constant (the Newton-Einstein gravitational constant) is used to obtain the equations of the gravitational field and torsion, the spin-torsion interaction turns out to be proportional to this constant and therefore extremely weak. In addition, the weakness of this interaction is repeatedly `aggravated' by the Planck constant, which in the classical version of the theory is linked with the spin value of the gravity source. Also much  smaller are the effects of “torsion repulsion”, proportional to the square of the torsion components.

In the Einstein-Cartan theory, generalized in spaces with affine connection, the components of the torsion tensor in the simplest case are included in the equations of the gravitational field as a sum of quadratic constants; these constant terms, in a certain sense, replace the Einstein cosmological term, which ensures the presence of some repulsive forces. The overall effect is that at a certain value of its three-dimensional radius (depending on the magnitude of the torsion components), the model of a closed FRWL universe stops collapsing into a singularity and begins to re-expand, which allows us to associate the presence of a torsion variable with the presence of repulsive forces opposing gravitational attraction. This constitutes a new approach to the singularity problem.

The resonance from the discovery of the \emph{torsion elimination of the cosmological singularity} among theorists in the 1970s  was so great that within a short time hundreds of works appeared developing the gravitational theory of the spin-torsion interaction. The most popular among others was the so-called Einstein-Cartan-Kibble-Sciama theory in which the equations of the gravitational and torsion fields follow from the simplest Lagrangian, the density of the scalar curvature of space-time with the metric affine connection (cf. \cite{hehl1,hehl2}). However, the general dissatisfaction with the limited possibilities of torsion in the Einstein-Cartan theory led to the appearance in the 1970-80s of many versions of torsion theories, in which torsion could propagate in a vacuum and was not always associated with the spin of the gravity source. Many of these theories were no longer purely `gravitational', since, in addition to the gravitational constant, new coupling constants were introduced into the corresponding lagrangians.

Besides, numerous attempts were made to include geometric torsion components in the equations of electrodynamics (e.g., \cite{sab2}), and as well to link hypothetical spin-torsion effects with the properties of vacuum and the evolution of individual astrophysical objects (e.g., \cite{sab1}-\cite{yef1}).

The ideas of spin-torsion coupling can still be traced in current publications, and indeed in the last 10 years or so there has been an explosion of interest in extensions of the original torsion gravity theories under the name of \emph{$f(T)$ gravity} (see \cite{clifton}, Sec. 3.5.1, \cite{sari1} for recent reviews). These theories have the advantage of having equations of second order, and they may account for dark energy and the late acceleration \cite{to1,to2}, but their relation to general relativity is not yet clear \cite{cr1,yang1}.


\subsection{Singularity theorems}
The definition of a singularity in cosmology as a `place' where infinities appear has the difficulty that it may be made to disappear by some kind of `surgery', removing the `bad' part and ending up with a spacetime without singularities. The existence of singularities was not clearly proved until the pioneering work of C.W. Misner \cite{mis4}, R. Penrose, S. W. Hawking, R. Geroch,  and others in the period 1963-1970 came along.

Penrose in 1965  \cite{pen} was the first to establish a black hole \emph{singularity theorem} to the effect that given certain plausible assumptions about the validity of general relativity, the structure of spacetime,  and the properties of matter fields, spacetime would be geodesically incomplete. Shortly after, Hawking adopted the topological methods introduced by Penrose and he proved several theorems for cosmological spacetimes \cite{haw}, eventually leading to an `all-purpose' singularity theorem in 1970 proved jointly by Hawking and Penrose \cite{hp}.

These theorems, together with related work by Geroch \cite{ge1,ge2} (see also the lectures \cite{ge3,ellis,penlect}, and the standard treatise \cite{he}) demonstrated once and for all the supremacy of the use of topological methods in studying the global structure of spacetime. They also set a firm basis for the consideration of \emph{the singularity problem}, which as emphasized by L. D. Landau, is one of the most important unsolved problems in theoretical physics \cite{khalatnikov,belinski}.

\subsection{BKL conjecture}
During the same period V. A. Belinski together with  E. M. Lifshitz and I. M. Khalatnikov (BKL) came up with a general scheme about the asymptotic  nature of spacetime and fields as one approaches the predicted generic singularity of the singularity theorems. This is a second aspect of the singularity problem, \emph{the nature of the cosmological singularities}. Their landmark work led to the formulation of the \emph{BKL conjecture}, a statement about the behaviour near the generic cosmological singularity. The resulting BKL behaviour  is essentially of a new and eminently complicated type. No power-law behaviour may be ascribed to it, and at \emph{each spatial point} on approach to the past singularity we have a situation described using a generalized Kasner solution and a subsequent homogeneous `Bianchi-type IX' oscillatory behaviour,  having an endless sequence of Kasner epochs grouped into eras all the way back to the beginning, cf. \cite{bkl2}. Misner \cite{misner} independently  discovered this evolution in hamiltonian variables in the homogeneous case, and coined the resulting cosmology \emph{the Mixmaster universe}.

Now in the general inhomogeneous case, the BKL conjecture has it, that  each spatial point, although evolving like one of these separate Bianchi IX homogeneous universes,   experiences the collective gravitational wave perturbations of neighboring spatial points, and the generic behaviour  is such that  almost all solutions of the Einstein equations approach an initial spacelike, vacuum-dominated, local, and oscillatory singularity, \cite{bkl3}.

However, in the case of inclusion of matter fields, in addition to the oscillatory behaviour of the solutions predicted by the BKL conjecture, there is a milder, monotonic approach to the big bang singularity, a type of  local Kasner regime,  namely, the asymptotic velocity term dominated (AVTD) behaviour, which is absent in the vacuum case. In this behaviour,  no terms can grow exponentially and spatial derivative terms become negligible asymptotically, for instance as in the case of a scalar field, \cite{bkl4}.

Although the BKL conjecture is unproved until today there are various different approaches to it. One approach to the general problem rewrites the system of equations into a form that directly generalizes from the spatially homogeneous case, and formulates the BKl conjecture in this framework, cf.  \cite{uvwe}. Another approach uses analytical and numerical techniques to support the validity of the BKL conjecture \cite{berger1}-\cite{berger5}. Still another approach to this basic problem is using new hamiltonian variables to reformulate the conjecture in a language suitable for M-theoretic cosmology, see Section \ref{m theory}.

\subsection{Chaotic cosmology and horizons}
One is therefore confronted with the general question of whether our universe evolves from a generic, possibly chaotic, initial state to the present one  characterized by a large-scale order and homogeneity, or the other way around - from an orderly initial state  to `chaos'.

Misner's  \emph{chaotic cosmology program} \cite{mis1,mis2} is based on the hamiltonian approach to the field equations of general relativity \cite{adm}. He thought to show that quite independently of any initial conditions in the early universe, more precisely if the universe was originally Mixmaster,  the long-term future evolution  will generically be characterized by a smooth and isotropic late universe, like the one we observe around us.  Misner's noble hope was that due to processes  like of frictional dissipation, irregularities in the initial distribution would have been  erased very early, leaving an isotropic universe: `...the emphasis was on the refocusing of cosmological theory from measuring the FRWL constants to explaining why we live in an FRWL universe...' \cite{mis3}. For simple Bianchi types this is indeed true, but later it was shown that such smoothing processes were probably not efficient enough \cite{ba-ma}.

In fact, the problem of explaining why the universe is homogeneous on the largest scales must be decided long before the universe became Friedmann-like,  due to a geometrical property  of the non-intersection of the past light cones called \emph{the horizon problem}. The existence of particle horizons would prevent the further synchronization of conditions in such a universe because domains in the microwave sky separated by more than a few degrees are causally disconnected, and, in fact, were so since the time of emission (last scattering surface). The Mixmaster universe, although describing a period before the Friedmannian stage of evolution of the universe, cannot solve the horizon problem and allow causal communication between such regions. This follows from detailed calculations of D. M. Chitre  (cf. \cite{chitre}, chap. V), resulting in the probability for the vanishing of a horizon in the model to be $0.02\%$.

However, the Mixmaster universe in the Misner or the BKL formulations is clearly a chaotic system as it is demonstrated by a combination of numerical and analytic evidence \cite{ba82,ba-co89,co-dem93,ho91,be94,dema} and refs. therein. This property is what makes the investigation of the generic inhomogeneous Mixmaster universe so difficult.
Nowadays, the Mixmaster universe is studied using qualitative methods and the theory of attractors \cite{bogo1,we,ring}.

Therefore the original hopes for explanation of the large-scale homogeneity through the homogeneous Mixmaster model did not work. Currently, the opposite direction is often cited as  popular,  the `order leading to chaos' point of view is generally preferred, cf. \cite{pee}, p. 212. A concrete suggestion for a `quiescent cosmology', that is a universe  with an isotropic initial singularity instead of a chaotic initial state, was made already in 1978 by J. D. Barrow in \cite{ba78}, and subsequently studied in detailed by many workers in the field, cf. e.g.,  \cite{co-ha,goode,scott1}, and refs therein.

\subsection{The finite-action conjecture}
The studies related to possible modifications of general relativity in conjunction with the works on the existence and nature of cosmological singularities in the 1960's led to doubts as to the general validity many of the previous conclusions in cosmological models, and in fact to a host of new and unexplored directions of research.

A unique characteristic of these early developments was the introduction of techniques from global differential geometry, topology and dynamical systems into the field of general relativity.
The problem of the \emph{existence} of cosmological singularities was adequately solved by the singularity theorems in general relativity, but \emph{the nature of singularities} remains a central problem in relativistic cosmology until today.

However, many are convinced that the predicted singularities in all important cases are accompanied by infinities in the curvatures and the thermodynamic properties of matter, and therefore at the singularities general relativity breaks down beyond repair. Thus many feel that the proved  existence of spacetime singularities in general relativity does not  really point to a new and worthwhile physical effect that could be fruitfully further explored within general relativity theory (e.g., `does the singularity resemble something like a hydrodynamical  shock wave rather than something more serious?' - cf. \cite{christo}).

Rather, the consensus seems to be that there is  new physics operating at a level  beyond general relativity and the standard model of cosmology, cf. e.g., \cite{he, pen-road,witten}. Then the singularity problem becomes only part of a more general (and complex!) \emph{issue of the initial conditions} for the early universe. This is so because one of the assumptions leading to the existence of singularities is that of the general validity of general relativity, and that assumption becomes  invalid when considered in the framework of possible \emph{geometric extensions of general relativity}, or of the still unknown theory \emph{quantum gravity} or, more generally, that of the still unfinished \emph{unified theory of all fundamental interactions}. In that sense, one may be allowed or even required to consider general approaches of the cosmological problem in all of these contexts, and this opens up a huge field of further investigation.

However, it is important to point out that the situation may not be so clear-cut as one would naively expect. This is so because there is a correlation between the lack of spacetime singularities in the curvature in general relativity or some future fundamental theory, and singularities that will by necessity appear in the universal action of the theory. Indeed the \emph{finite-action conjecture} \cite{bati88},
\cite{bafinite}, \cite{le-ste} relates to exactly that: The finiteness of the universal  action of some theory applicable to cosmology depends on the existence of past and future spacetime curvature singularities because that finiteness results from an integration over  finite time intervals.  When action singularities are avoided then new spacetime singularities will arise and \emph{vice versa}, when spacetime singularities are absent in a theory then new action singularities will emerge. Another important point is that since paths of finite actions have generally zero-measure contributions to the path integral for many examples such as the simple harmonic oscillator \cite{coleman0}, the correct path integral in cosmology must be such that  such finite action contributions become the dominant ones, and this might be a very restrictive requirement. This is a frontier of modern cosmological research.

\section{Third period, 1980-2000}
In this Section, we discuss the following novel ideas that emerged in this period:
\begin{itemize}
\item Inflationary cosmology
\item The multiverse
\item Wave function of the universe
\item The cosmological measure problem
\item Baby universes and wormholes
\item Kaluza-Klein  universes
\item String cosmology
\item $f(R)$ gravity and cosmology
\item The issue of cosmological stability
\end{itemize}
\subsection{Inflation}
In the late 1970s, dramatic discoveries in particle physics like asymptotic freedom and the various issues and approaches of  how vacuum fluctuations may lead to models of grand unified theories of electromagnetic, weak and strong interactions, completely changed that field, and it was not long before the consequences of these developments for `early universe' cosmology were followed. Particle physicists found in cosmology a testing ground for their theories, while cosmologists realized the power of high-energy arguments for the complex problems that they were facing with at the time. \emph{Particle cosmology} was thus born because of the need to better understand the behaviour of matter and fields in the extreme conditions of the big bang.

\emph{Inflationary cosmology} \cite{star0,kaz,guth1}, a paradigmatic application of particle physics ideas to cosmology. Unlike more ground-to-earth applications of nuclear physics and elementary particle theory to cosmology, like big-bang nucleosynthesis and baryogenesis, the appearance of inflation was a truly exotic idea, but at the same time had some attractiveness to many cosmologists who had a good background of general relativity. It was based on the transient effects of a scalar field for an accelerated expansion of the universe (in de Sitter's and in Lema\^{i}tre's models discussed earlier,  acceleration appears as a \emph{permanent} property, not a transient one).

At a proper time hypersurface $t\sim 10^{-35}sec$, we imagine that the particles comprising the material content and occupying different vacua of different energies were displaced from their equilibrium states and were able to move to other vacua of lower energies. During this motion between  initial and final vacuum states occupied by matter, the vacuum energy liberated creates a gravitational repulsive force similar to the original Einstein's cosmological constant and so the universe accelerates for the brief period between the two, the old and the new,  vacuum states.

This is very important. For each such pair of vacua, particles arriving at  new vacua are able to form  `bubbles of new vacua' nucleating at different spatial regions in different times. According to Guth, this phase transition supercools the universe by 28 or more orders of magnitude below a critical temperature, and  then leads to a huge expansion of exponential growth accompanied by a disastrous process of all sorts of irregularities in the curvatures and the matter density arising in this phase of evolution of the universe. (This was ameliorated, however, in subsequent models by other authors making the motion between vacua \emph{slower}-leading to a unique bubble comprising the visible universe.) In fact, a slight change in the original scenario of Guth's showed that  in the `new inflation' scenario (as in various others-see below), inflation is in fact eternal \cite{stein0}.

But the inflationary idea, common to all inflationary models that followed Guth's,  also leads to a much larger, homogeneous and isotropic universe, with many of the puzzling issues of the previous models practically solved. That includes the horizon, flatness, monopole, as well as other perplexing issues, unexplained in the standard model of cosmology. For instance, the visible universe today - our own bubble -  comprised a huge number of causally disconnected regions at Planck time (and also at the time of inflation). But during inflation one of these regions was inflated to encompass all others, allowing thus a restored type of causal communication between them and leading to today's observed uniformity.

In fact, it is a most important property of the resulting inflationary picture of the early universe, also improved and generalized by various later modifications or extensions, that by the same mechanism it also   explains other additional properties:  the temperature and energy-density fluctuations of the cosmic microwave background radiation as a result of the quantum phase during the inflationary stage, namely, the quantum fluctuations of the scalar field driving inflation  (see \cite{weinberg1} for background in the theory of cosmological fluctuations, \cite{bran0} for an early review of the inflationary approach to this problem, \cite{kolb} for the situation in early 1990's, and \cite{weinberg2}  for a modern review of this idea).

Today the development of the inflationary picture of the early universe continues. There are practically hundreds of model-dependent implementations of inflation in different theories, for instance in  relativity, modified gravity, supergravity, superstring cosmology, M-theoretic cosmology, etc, but there are two issues of importance related to the inflationary scenario that require attention, and which show that inflation is certainly not the last word about the structure of the early universe.

The first is to test how successful inflation models fit  cosmological data coming from  \emph{observations} made by various satellites and instruments,  COBE, PLANCK, WMAP BICEP/KECK, etc. Indeed, many of the most popular inflationary models are now definitely falsified by these data. This is made possible by the consideration of the slow-roll parameters $\epsilon$ and $\eta$ defined by the scalar field potential and its first and second derivatives. Eventually the predictions of inflationary models  are measured by certain combinations of the slow-roll parameters calculated in the models, the primordial tilt $n_s$ and the tensor-to-scalar ratio $r$. These are functions of the number $N$ of the inflationary $e$-folds which occur between the time when a given perturbation leaves the horizon and the end of the inflationary period,  and describe quantum fluctuations induced by inflation such as  the gravitational wave energy spectrum. Interestingly, current observations give a value of $r<0.036$ at $95\%$ confidence \cite{bicep}, too low for most inflationary models to predict.

The second issue is what would be a believable \emph{prequel to the inflationary stage} that would make the inflationary idea more natural. We shall come back to this  issue in subsequent Sections of this paper.

\subsection{Multiverse}
Even though according to inflation the bubble leading to our visible domain adequately expanded and inflated solving in this way many of the cosmological conundrums of standard cosmology, the  complex process of creation and evolution of the whole network of such distinct  inflating bubbles  may lead to a very awkward situation, now elaborated to what is called the \emph{multiverse}.

The fact that inflation can be \emph{eternal}  was realized soon after Guth's original idea, by Paul Steinhardt \cite{stein0} and  by Alexander Vilenkin \cite{vil0} for new inflation, and later by Andrei Linde  \cite{linde1} in the context of the chaotic inflation scenario, where the potential of the scalar field driving inflation has no flat plateaux. Linde used heuristic arguments to imply that the  process of large-scale fluctuations of the scalar field leads to an \emph{eternal, self-reproducing, inflationary multiverse}. In this, each inflating bubble consists of further inflating subregions with each one these in turn having further similar ones,  at infinitum,  a process  seemingly progressing endlessly to the future as well to the past. In this multiverse, the density parameter $\Omega$ varies continuously between 0 and 1, so our bubble together with infinitely others with $\Omega\in(0.2,0.3)$ lies somewhere in the structure with nonzero probability.

It is a difficult problem how to calculate probabilities in the multiverse \cite{linde95}, and in a situation like this one may produce arguments leading to an overall controversial picture (cf. e.g., \cite{vil3} and refs. therein) for the probability of the occurrence of a stage of inflation. For instance, in \cite{linde02}, \cite{wald-hollands} one finds opposite views and probabilities ranging from almost one to almost zero.  Despite being hard to believe,  this picture appears nevertheless as a natural extension within the wider context and philosophy of inflationary cosmology, and some recent progress has indeed been made (see Section \ref{landscape}). In fact, most authors agree that inflation is future-eternal \cite{linde2},  but  probably not being so in the past \cite{guth2} without violating the null energy condition \cite{ag}.

\subsection{Quantum cosmology}
For the consideration of the  \emph{initial state of the universe} as a quantum problem,  the so-called program of  \emph{quantum cosmology}, one such approach is through the \emph{Wheeler-DeWitt equation} advocated by B. DeWitt in  \cite{DeWitt} and by J.A. Wheeler in \cite{wheeler},  describing the evolution of the quantum wave function  of the universe considered as a quantum object. Solutions of the Wheeler-DeWitt equation then give probabilities for the universe to evolve from one state to another, provided one gives prescribes how to specify `initially' both the probability amplitude and its first derivative normal to the initial hypersurface in superspace (this is a set describing the space of all 3-geometries). One approach to the study of this equation was to consider the limiting case of `minisuperspace', a subset of the full problem where most of the gravitational degrees of freedom are frozen out. This turns the whole quantum cosmology problem, considered in `canonical quantum gravity', into a more manageable quantum mechanics problem, and many different models have been considered  in this framework.

A prequel theory to inflation remains a very mysterious problem and its solution largely unknown even today. This  is perhaps one of the most important ingredients necessary to make our current(and future) cosmological standard models more complete. In a sense, initial or boundary conditions are necessary for any cosmological model to be more predictive. All these issues became more substantial and received a blow of original ideas in the 1980's (the decade 1980-90 may be called indeed a `golden decade' in theoretical cosmology research).

In 1983, J. Hartle and S. W. Hawking introduced a radical approach using path integral methods to describe the problem of initial conditions in quantum cosmology introduced in the 1960's \cite{hh}. That approach called the \emph{no-boundary proposal} for the wave function of the universe suggested a beginning of the universe  not as dramatic as that appearing in the standard model of the universe. In fact, there is a beginning but the initial singularity is now replaced by a situation where time is imaginary, there is a smooth passage to a quantum regime, a creation `from nothing'. The \emph{Hartle-Hawking wave function} also peaks as most probably states those that describe infinitely large and empty  universes.

However, it soon became clear that other, equally probable and similarly constructed, `creation-from-nothing' proposals are possible, but with completely different predictions. For instance, a small, hot early universe, pretty much like the standard big bang model of the classical universe is the prediction of \emph{A. Vilenkin's tunneling model} creation-from-nothing 1985 model \cite{vil}, that also predicts a period of exponential expansion \cite{vil2}.

Soon however, the problem arose of what it means for the universe to be in a \emph{typical quantum state}, and what is the meaning probability in quantum cosmology as well as the possible set  of all these different wave functions, (see e.g.,  \cite{gib}  for a clear discussion of the difficulties in defining a  probability of inflation in a simple minisuperspace model). This is a very deep problem and is related to the measure problem in cosmology and the issue of how to define probabilities in a cosmological setting (see next Section).

The first works resulted in a huge proliferation of very interesting subsequent papers on quantum cosmology during the 1980's (see \cite{hall} for a bibliography of papers for that period), addressing an impressive variety of very deep questions and meaning in cosmology (prediction, time, creation, etc).

Finally, we mention a very different approach to quantum cosmology, \emph{loop quantum cosmology}, LQC. This is based on an approach to the quantization of gravity using new hamiltonian variables (the Ashtekar variables), and taking seriously the Dirac quantization method of field theory. LQC also uses a different set of connections than the Levi-Civita one, the spin connection, and this results in an improved behaviour of the original Wheeler-De Witt equation. This also gives a bounce instead of the initial singularity in FRWL models considered in LQC, whereas many other cosmological properties acquire a new light in this framework. For reviews, see \cite{ash1,bojo}.

\subsection{Measure-theoretic cosmology}
There are many situations in theoretical and mathematical cosmology where one is interested in `typical behaviour', the `most probable state' a cosmological system may assume, in short, one is interested in a measure of the `degree of genericity of a given property'. For example, in inflationary cosmology, there is the question of how typical is an inflationary stage within a given set of solutions. For instance, consider a flat FRWL model with a perfect fluid with equation of state $p=w\rho$, where $p, \rho$ are the pressure and the density of the fluid respectively. Then for different values of the parameter $w$, one obtains different dynamical regimes described by exact solutions for given $w$ values, and the question arises as to what is the long-time behaviour of the solutions of the system. For a scalar field  potential quadratic in the field, it may be shown \cite{bgzk} that inflationary stages are  an unavoidable property of most solutions of the associated dynamical system, and in addition for exponential potentials, power-law inflationary stages are likewise a stable attractor of the system \cite{hall2}.

To study probabilities in a classical or quantum context in cosmology more generally, it is necessary to have a suitable measure in the space of all classical solutions of the theory, much like the Liouville measure (i.e., the volume element) in statistical physics.

Such a measure was first proposed by M. Henneaux in \cite{h1} and by G. W. Gibbons, S. W. Hawking, and J. M. Stewart in \cite{ghs}, and we shall refer to it as the \emph{HGHS canonical measure} on the set of all universes. This measure plays the role of a Liouville-type  measure in a minisuperspace approximation of the cosmological phase space, and its construction assumes a hamiltonian flow and associated symplectic structure and constitutes an adaptation of the symplectic quotient construction in  cosmological problems.

The application of the HGHS measure to cosmology, most importantly perhaps to the structure of the multiverse,  is, however, plagued with various infinities. In addition, there are various conflicting claims in the literature for inflationary vs. noninflationary solutions,  in many cases of interest such as a massive scalar field \cite{hpage}, $R^2$ inflation \cite{p1}, Bianchi-I inflation \cite{cp}, the HGHS measure gives infinite answers on both types of solutions. In \cite{gt} (see also \cite{p2}), however, a cutoff in the dynamics is introduced for which the resulting measure becomes finite because the probability of $N$ $e$-folds of inflation is then exponentially suppressed, thus making inflation improbable.

Other studies of the measure such as \cite{schi} point out related difficulties associated with the HGHS measure, such as the use of minisuperspace approximations, the lack of inhomogeneity considerations, as well as the lack of any interaction between different pocket universes in the ensemble to which the measure is applied (so as to allow for `equilibration' on the relevant scales), while a comparison between different measures can be found in \cite{lin-noor}. A broader perspective on measures in the multiverse is found in \cite{emm}.

Despite all these extremely valuable works, it is indeed true that the measure problem is currently unresolved and more research is clearly needed in this most interesting problem of mathematical cosmology.

\subsection{Wormholes and baby universes}
Wormholes connect parts of an asymptotically flat (Euclidean) region in spacetime, or perhaps two different asymptotically flat regions, and are important because according to S. W. Hawking  describe possible quantum states of closed universes that branch off from our own \cite{haw-worm}. One may imagine wormholes as small tubes connecting various types of two closed 3-boundaries, or `baby universes', the latter being closed carry zero energy and momenta. (The word `semi-wormhole' is sometime used to designate a tube that connects a state with no baby universe to another with one.)

The importance of wormhole-like objects in quantum cosmology took off in the late 1980's when S.   Coleman showed how they can be used to make the cosmological constant vanish \cite{coleman1}. For this purpose, he used the Hartle-Hawking wave function to show that it had a peak at configurations for which the cosmological constant vanished. In this situation, $\lambda$ is contained in the leading term in an expansion of the classical action, while all other higher-order terms (i.e., $-R+aR^2+b\textrm{Ric}^2+c\textrm{Riem}^2 +\cdots$, with the coefficients depending on the wormhole being summed) vanish, being  the saddle points of the action, i.e., Einstein spaces. Then the effective action $\Gamma$ is found to be simply inversely proportional to $\lambda$, hence peaked at $\lambda=0$.

This spectacular result was criticized because it was thought that it led to a situation having a  network of large wormholes on very large scales with very high density (of the wormhole ends  in the dominant Euclidean field configurations) - a catastrophic prediction of strong, nonlocal interactions \cite{fs}.  Nevertheless, it led to a very large number of works on the varied  physical effects of wormholes (see, for instance, the interesting proceedings at that time \cite{baby-proc}).

Another cosmological application of wormholes is in the construction of bouncing models which avoid an initial singularity. It was first shown by by M. S. Morris and K. P. Thorne that wormholes can be solutions to the Einstein equations that violate the null energy condition but allow for time travel \cite{mt1,mt2}. Various  such wormhole solutions exist with matter models, or various scalar fields, in modified gravity,  scalar-tensor theories, or in brane universes, see \cite{bro1,bro2}. A particularly interesting wormhole solution which requires no violation of the energy conditions, is a cylindrical wormhole with spherical topology near the throat \cite{bro3}. Another example of such a configuration is a generalization of the Tolman universe of Section \ref{tolman}. That was  bouncing universe but the nature of the bounce is somewhat ambiguous as it is not precisely specified. Wormhole solutions connecting a previous phase in the evolution of the universe to the current one can  provide the missing link, and many such models have been built (see \cite{visser} for a textbook presentation). A Tolman wormhole is an example of such a  bouncing model with the strong energy condition being violated \cite{cvisser}.

\subsection{Kaluza-Klein cosmology}
An additional, important input to `alternative' cosmology was provided by the multitude of possibilities in the development of \emph{Kaluza-Klein cosmologies} (see e.g., \cite{freu,abb,ok},  \cite{kolb} for a consice, almost textbook treatment, and \cite{wesson}  more suitable for relativists (this paper contains almost 400 references)).

In such a higher-dimensional setting, there are two kinds of spatial dimensions, the usual `external' (large) ones,  and the internal (small, that is `compact' ones, see below) dimensions which play a key role in determining the structure of the physical laws in the overall higher-dimensional setup (extra timelike dimensions are associated with ghost instabilities and are thus less favorable). There is a fundamental `cylinder condition' which defines circle-compactification, meaning that nothing depends on the extra $Y$ coordinate, or equivalently, the space that is defined by the extra coordinate(s) is compact in the subspace topology.

This leads to a very small scale (volume) of the compact dimensions compared to the remaining ones, and so their associated energy must be very high. Since in these models, the Newton constant, which gives the range of the gravitational interaction,  appears as inversely proportional to the very small volume of the compact internal dimensions, this led some to propose a higher-dimensional mechanism for explaining why the gravity scale is so much larger than those of all the other interactions, the so-called hierarchy problem  (see Section \ref{brane}).

Also the fundamental physical constants that appear in the low-energy theory that we observe have values that crucially depend on integrating over the extra dimensions. The internal dimensions are also static, in distinction to the external ones that are dynamic, usually expanding or contracting,  and this situation has led to a great number of works that approach the standard problems of cosmology (nature of singularities, bouncing models, horizon problem, ...) from such a higher-dimensional perspective (cf. the quoted references for various examples of such models).

One particularly interesting development in this context has been the issue of whether or not chaotic Mixmaster behaviour is possible with more spatial dimensions. After, initial results \cite{furu}, \cite{ba-stein} about the disappearance of chaotic behaviour in vacuum 5- and 7- dimensional Kaluza-Klein universes respectively with static internal dimensions,  in a series of very interesting papers in the middle 1980's, \emph{J. Demaret, et al} argued  that in homogeneous \cite{dem1} as well as inhomogeneous \cite{dem2} vacuum Kaluza-Klein cosmologies more dimensions lead to less chaos, in particular, chaos disappears in eleven or more spacetime dimensions, but it remains  in homogeneous vacuum  Kaluza-Klein universes in spacetime  dimensions between 4 and 10 \cite{dem3} (the spacetime dimension 11 has been  called the `critical' dimension for chaotic cosmology in \cite{hosoya}).

But two issues in all such models remained: What is that which makes some dimensions expanding while others are kept small, and secondly how does one make the (small) size of the internal dimensions stable with respect to physically relevant perturbations? Both these problems are addressed in more elaborate extensions of the interesting initial Kaluza-Klein cosmologies of the 1980's.

\subsection{String cosmology}\label{string}
A new chapter in mathematical and theoretical cosmology opened with the application and exploitation of the new duality symmetries between  string-theoretic models  and 11-dimensional supergravity, suggesting the existence of a still larger theory - M-theory - in which gravity being supersymmetric propagates in the 11th dimension while the remaining interactions are constrained on suitable 10-dimensional hypersurfaces. This new setup requires  in addition to the metric, other fields - for instance two more massless states, the dilaton and an antisymmetric tensor field (commonly called `the axion').

For cosmology, all this new information translates into a huge variety of suitable gravitational lagrangians including all sorts of other fields, leading  to solutions which cannot exist in general relativistic or other similar contexts. The result is \emph{string and M-theoretic cosmology}, a new a largely unexplored field of investigation (for  excellent textbook introductions, see \cite{gasp}, and the last Part of \cite{peter}.

The duality symmetries when applied to a homogeneous Bianchi I cosmology lead to the so-called scale-factor dualities, which  not only invert the scale factors but at the same time shift the dilaton, therefore making this a purely string-related effect. When combined to time-reversal symmetries, and applied to a Friedmannian background, one is led to a good result: In \emph{the pre-big-bang model} of M. Gasperini and G. Veneziano \cite{gasp-ven}, the singularity has moved to infinity, and its place is taken by a smooth evolution bouncing through it from a previous expanding  to a post big bang contracting homogeneous and isotropic state. (More general results are possible in anisotropic contexts.)

The resulting universe-model is characteristically distinct from the hot big bang cosmology, in that the curvature has a maximum at the end of an initial cold, unstable, and vacuum state. One may compare the quantum cosmological approaches to the hot big bang vs. the string perturbative vacuum state of string cosmology in a quantum setting, where the latter overpasses some of the well-known problems of the standard approach to quantum cosmology, and introduces a tunnelling not `from nothing' but from the initial string state (see \cite{gasp} for a complete treatment of the various results in this approach). A potential issue with this model is, however, the possible instability of single semi-classical trajectories, an issue which could be bypassed in a path integral approach.

More generally, string cosmological models offer new possibilities for cyclic behaviour (cf. the interesting qualitative work \cite{coley1} and refs. therein),  exact solutions of Bianchi type made possible with new forms of the antisymmetric $H$-field \cite{ba-kun1}, and a  variety of inhomogeneous solutions with an initial Kasner era \cite{ba-kun2}. For more results of this type, see the comprehensive review \cite{wands}.

\subsection{$f(R)$-cosmology}
A rather different modification of the basic dynamical form of general relativity than the Brans-Dicke or more generally the scalar-tensor theories, - higher-order gravity - appeared very early and continues in new forms until today, having major ties with early universe cosmology. Based on earlier impressive and original work of C. Lanczos (cf. \cite{lancz} and refs. therein) and others on the influence of higher-order invariants of the Riemann tensor on the structure of general relativity, a particular choice eventually seemed simple and general enough: the gravitational lagrangian be an arbitrary analytic function of the scalar curvature, $f(R)$. In fact, this leads under a metric variation to fourth-order field equations, not second order like general relativity (while other types of variation may lead to second order ones, see below). The result is a new theory of gravitation, \emph{$f(R)$-gravity}.

Keeping $f(R)$ analytic means that this type of theory  will be expected to play a significant role in the extreme conditions of the \emph{early} universe where higher-powers of the scalar curvature become important, but not for late-time evolution (where such powers become negligible).
$f(R)$-gravity is hence expected to convey some of the importance of the influence of quantum gravitational effects in the early universe even though everything is classical.

Some first impressions of $f(R)$-gravity to cosmology were given in the period considered (see e.g., \cite{buchdahl,gu1,nariai,kerner}), but the field was meant to blow up in activity starting in the mid 1980's, and this continues until the present time (see the many reviews of this theory towards the end of this paper).

In 1983 in a seminal paper, J. D. Barrow and A. Ottewill \cite{ba-ot} showed that under especially  simple algebraic conditions, the whole family of theories in $f(R)$-gravity allows for inflation, as well as it provides  stable FRWL, de Sitter  cosmologies with respect to perturbations outside general relativity, that is in $f(R)$  theories. In addition, it was shown a little later that $f(R)$-gravity is conformally equivalent to Einstein's theory when another purely cosmological, self-interacting scalar field is added \cite{ba-co1}\footnote{In distinction to the case of the BD, or scalar-tensor, theory, this scalar field was defined as proportional to the $\ln f'(R)$. Although there are claims in the literature of a relation of $f(R)$-gravity to Brans-Dicke theory by pure redefinition of the fields, these are based on non-cosmological scalar fields (such as proportional to the scalar curvature thus becoming negligible at cosmological scales), and extra conditions like $f''(R)\neq 0$, which unnecessarily restrict the original theory. On the other hand, conformal equivalence does not require such relations between the two classes of theory, and shows that both $f(R)$-gravity and Brans-Dicke type  theories are \emph{directly} conformally equivalent to GR plus suitable scalar fields.}.
This equivalence of the two dynamically different theories required the same type of matter as inflation did, but it also showed that $f(R)$-gravity leads to a symmetric-hyperbolic system, in contradistinction to other types of modified gravity. This equivalence also allowed to transfer a number of results from general relativity to the generalized framework of $f(R)$-gravity, including the singularity theorems (because conformal transformations respect causality).

With regard to conformal transformations in such contexts, there is an important geometric generalization of general relativity, conformal (Weyl) gravity. This theory has a unique place among all theories quadratic in the curvature invariants, since it is conformally invariant. (It is also closely related through  the Gauss-Bonnet theorem to a particular Bach lagrangian.) This leads to intriguing properties and connections with various important and largely unexplored issues, such as what breaks conformal invariance, and how this is related to a mumber of cosmological puzzles. For more on this theory cf. \cite{kaz-mann}.

The development of \emph{$f(R)$-cosmology} continues today in an accelerating pace. Two developments are of particular importance. The alternative Palatini formulation of the $f(R)$-gravity equations \cite{cmq} leads to a reduction of order and can therefore lead to simplified treatments of a number of issues in cosmology (for a review of cosmology in Palatini theories, see several of the reviews cited at the end of this paper).

Secondly, there are new developments of \emph{no-scale} $f(R)$-theory \cite{ba-coTR}, that is lagrangian $f(R)$ theories in which the field equations are traceless versions of the standard ones. An important feature of these theories is that conformally they become GR plus a scalar field but with the crucial difference that the self-interacting potential is scale-invariant. This exact result implies different CMB parameter forms for all previous forms of $f(R)$-gravity. For the traceless version of the quadratic theory, the prediction for the  tensor-to-scalar ratio is given by, $r=12/(b^2 N^2)$, with  $b\neq 1$ is the new condition necessarily required in the traceless version \cite{ba-coTR}. (Here $N$ stands for the $e$-fold function that measures the length of inflation, while $b$ is the crucial new arbitrary constant related to the scale invariance of the potential when we scale the field in traceless theory.)  This theory accommodates any $r$-value however small.

\subsection{Cosmological stability}
In Section \ref{static}, we mentioned that the most important property of the Einstein static universe was its transient character. This is based on its instability, as in the Eddington-Lema\^{i}tre cosmology. In fact, the issue of the instability of the Einstein static universe is much wider, and has been investigated by different authors who proved instability in various current contexts.

Starting with a  radiation-filled Einstein universe, instability was shown in terms of inhomogeneous oscillatory modes \cite{harrison67}, or in a fluid-filled model with respect to conformal metric perturbations (where the interesting result is that there are regions of stability depending on the sound speed)  \cite{gibbons87}. Studies of instability were then exploited by G.F.R. Ellis and R. Maartens to introduce the \emph{emergent universe} \cite{ellis-maartens}, where in the context of inflationary cosmology, the Einstein static model serves as a prequel to the inflationary stage: an transient initial state with matter being a scalar field whose vacuum energy determines the cosmological constant of the model. This model has no singularity, no horizon problem, and most importantly does not need a quantum era in its early evolution, it is past eternal. Its instability of the initial Einstein static state  is proved \cite{tsagas}, but the effects of other, inhomogeneous, nonlinear perturbations are not yet concluded (see, however, a proof against Bianchi-IX homogeneous perturbations in \cite{barrow-yamamoto}). Last, the Einstein static universe, perhaps even as a transient model,  does not seem to pass the finite action requirement \cite{barrow20action}.

The issue of cosmological stability is generally related to various geometric properties of the flow of a system of equations. For Bianchi universes in vacuum or filled with various fluids,   exact solutions are known and so  issues of stability have been studied by many authors (see \cite{ew} for a review of the exact solutions in this context). The systems are described by ordinary differential equations but the stability of most types (e.g., those of type V, VI, VII)  becomes nontrivial to decide because of the typical appearance of zero eigenvalues in the case of future asymptotic evolution \cite{ba-so}.

For Bianchi cosmologies with a positive $\lambda$ there is a result of Wald \cite{wa83} to the effect that all (including B-IX with large enough $\lambda$) approach the de Sitter solution, and are therefore isotropize. There were various such results in the 1980's and the 1990's pointing to the validity of a \emph{cosmic no-hair theorem} for such spacetimes also in extended frameworks \cite{cf1} such as higher-order gravity \cite{cm1}, or quantum cosmology \cite{ym}.

Stability really depends on how one sets the problem, and there are various different and inequivalent definitions  in the literature. It is closely related to the problem of describing the asymptotic states of a cosmology, and for this there exist various different approaches (see, eg., \cite{wh,bogo} for homogeneous cosmologies). This is a vast field within mathematical cosmology, rapidly expanding today. For inhomogeneous models, the situation is far less well-understood, see next Section.

\section{Fourth period, 2000 until today}
During the last twenty years in the evolution of mathematical cosmology one sees an explosive mixture of joint developments in many of the aforementioned areas as well as the formation of many novel ones. It is fair to say that the whole field has matured to a degree beyond recognition as compared to, say, its image 50 years ago. In this last part of the development of the field, we shall focus primarily on the following topics:
\begin{itemize}
\item M-theory and cosmology
\item Braneworlds
\item The landscape
\item Topological issues and dynamical evolution
\item Genericity in cosmology
\item Models of dark energy

\end{itemize}

\subsection{M-theoretic cosmology}\label{m theory}
It was a surprise to many that the quantum consistency of superstrings eventually required the existence of further (than strings) low-dimensional objects, the $p$-branes, with $p$ taking `small integer values'. Once one allows for strings (that is $p=1$ objects) into the theory, however, it is completely natural to expect that other such objects play an important role and consider these  other possibilities. Branes are ubiquitous and largely unexplored, and given the various dualities between the original five string theories and 11-dimensional supergravity, it is only natural to explore them in this generalized context of interconnected theories coined \emph{M-theory}. In fact, in the \emph{string or brane gas scenario} of Brandenberger et al in 2000 \cite{bra1}, the universe is supposed to be a `hot soup' of branes of all $p$ dimensions, topologies, and spatial orientations, and all internal dimensions are assumed compact, in an attempt to explain why only three dimensions are surviving. In the simplest models of this kind,  branes of larger dimensionality have higher energies and annihilate first, leaving only strings. Brane cosmology is a very rich subject, and some aspects of it are treated in Section \ref{brane}.

A  most important aspect of M-theory cosmology is revealed in the study of chaotic Mixmaster behaviour in these frameworks. Initial results \cite{ba-da1} pointed to the disappearance of the BKL chaotic oscillations on approach to the string cosmological singularity in\emph{ homogeneous} Mixmaster models, due to the combined effect of both the dilaton and the axion fields. A more elaborate treatment of this problem brings us into the realm of the general effects of $p$-form fields often coming from supergravity theories and  M-theory.

A spectacular effect of these fields was shown to exist  in the generic inhomogeneous case of the old Mixmaster universe but now considered  in the context of superstring or M-theory cosmologies,  where it was shown that the role of the $p$-form fields is similar to that of the curvature terms in the Einstein theory, namely they act like potential walls, thus preventing the possibility of free Kasner motions of the universe point \cite{dam1}. The overall result in any spacetime dimension in the general inhomogeneous case is again the BKL behaviour, even in the case of 11 or more spacetime dimensions where that erratic behaviour was known to be absent in pure Einstein gravity \cite{dam2}. This led to the interesting speculation of a possible \emph{de-emergence} of a classical or even a quantum behaviour of spacetime as we approach the initial singularity in these models, and the possibility of an effective, purely algebraic description of the dynamical behaviour of the universe near the singularity  in terms of hyperbolic Kac-Moody algebras \cite{dam3}.

Today, about twenty years since its original conception, M-theoretic cosmology is only just beginning. In a series of interesting papers, A. Lucas et al \cite{wal1}, (see also \cite{wal2} and refs. therein),  introduced and studied a general framework for doing cosmology  in type I superstring and M-theory, with particular emphasis on the relation between supergravity  $p$-brane solutions and cosmology. Because of an allowed exchange between the time coordinate and the transverse spatial coordinate in these solutions, the intriguing possibility opens to explaining a number of cosmological phenomena through stringy considerations.

For  qualitative  applications of dynamical systems techniques in the context of an FRWL model with six Ricci flat internal dimensions and the 11th dimension compactified on a circle, see \cite{coley2}. A feature of M-theory cosmologies appears to be  their transient acceleration, see \cite{town1} or many related references, and \cite{and1} for a more elaborate analysis of this property. An interesting ingredient used in these works is a correspondence between the space of flat FRWL cosmologies and that of geodesics of a suitable target space, with accelerating ones occupying a special subset of the lightcone \cite{town2}.

For background necessary for string- and M-theoretic techniques particularly adapted for the relativist, see the relevant sections in \cite{gasp} as well as \cite{mar}.

\subsection{Braneworlds}\label{brane}
Braneworlds, or $p=3$-branes, represent universes (i.e., 4-dimensional spacetimes) lying inside a higher-dimensional space dictated by superstring theory, in which the three particle physics interactions `live on the brane' while gravity propagates freely in all dimensions of space, and is much weaker than the other forces. This is in sharp contrast to the usual Kaluza-Klein cosmology wherein the extra dimensions are necessarily static and presumably compactified,    hence unobservable.

Brane theory seriously elaborates further on this very original Kaluza-Klein idea. There is a set of consistency conditions of the quantum superstring theory that leads naturally to the exploration of the possibility of having `large extra dimensions', dimensions that have a compactification scale not at Planckian energy but much lower, possibly comparable to the currently favorable ones (cf. \cite{ant1,ant2} and references therein).  A concrete  realization of these ideas was originally introduced by Randall and Sundrum \cite{randall} in 2000, where the Einstein cosmological equations are modified by the addition of extra terms coming from the brane embedding in the bulk space, as well as the bulk space itself.

In the resulting setup  new possibilities are possible for inflation, gravity `induced' on the brane by localizing matter,  or more general cosmological dynamics, (see e.g.,  \cite{gasp}, chap. 10 for a short review). One may imagine a number of such lower-dimensional universes coexisting in the parent space and moving in one or more of the extra dimensions. Using the mathematics of submanifolds, it is not very difficult to describe the geometric properties of this setup as well as the modified Einstein equations, cf. \cite{maartens}, \cite{langlois}.

In fact, there are several scenarios in the literature about possible brane models of the early universe, a particularly interesting one being the \emph{ekpyrotic universe} proposed in 2001 by J. Khoury, B. Ovrut, P. Steinhardt, and N. Turok \cite{stein} (see \cite{lehners} for a review with many references). Here,  two parallel branes approach each other, collide, and then rebound, moving in one of the spatial dimensions available. The big bang occurs as a bounce periodically an infinite number of times, each time there is a collision, and then there is a cyclic process  of contraction, bounce, and expansion, like in the old Tolman oscillating universe. There is a stage of early acceleration in this model which keeps the entropy from diverging in the cycle-to-cycle evolution after many cycles, and according to the authors there are no infinities in the defining quantities like in the standard cosmological model. The ekpyrotic model is but one of many possible bouncing cosmological models (see \cite{bran2} for a related review).

Braneworld models nowadays abound. They have  different motivations than the original brane models,  but they all show the richness of  brane cosmological ideas. Two such models are briefly noted below, one motivated  by an attempt for a  successful resolution of the cosmological constant problem, and the other by an adaptation of holographic ideas in cosmology. For a resolution of the $\lambda$ problem in a higher-dimensional, brane setting, one uses a self-tuning mechanism and considers a single 3-brane universe in a 5-dimensional bulk space with  \emph{linear or nonlinear bulk-fluids} (cf. \cite{ant-co-kl} and related refs. therein). The main question here is  whether there are regular solutions that satisfy a number of plausible physical criteria (finite Planck mass,  energy conditions, etc). In these braneworld  models,  `everything depends on the extra transverse $Y$-coordinate', grossly violating the cylinder condition met in the original Kaluza-Klein universes. On the other hand, a particularly interesting feature of these models is the fact that a Wick-type of exchange between the time and the transverse spatial coordinates leads to  cosmological models with many features having brane-theoretic origins.

The last  brane model we discuss  is  \emph{ambient cosmology}. This is an attempt to relate holographic techniques \cite{witten} in a braneworld setting with methods of conformal geometry to motivate the idea that the brane representing `our world' has  moved to the conformal infinity of the bulk, the `ambient space'. The resulting \emph{ambient cosmology} has a number of novel characteristics that allow for previous problems such as the singularity problem and the question of cosmic censorship to be resolved very smoothly in this setup \cite{ant-co}.

In conclusion, braneworld cosmology  presents interesting, well-posed problems for the future  in terms of challenging the predictions of inflation as viable alternatives, as well as a variety of mathematical issues which, in face of the great number of unexplored possibilities in this part of cosmological model building, become particularly attractive to tackle.

\subsection{Measures in the landscape}\label{landscape}
Interestingly, the picture of the inflationary multiverse according to eternal inflation and developed in the period 1980-2000, was found to be strongly supported by a  string- or M-theoretic prediction of a huge collection of allowed, or even predicted, different vacua known as \emph{the landscape} \cite{suss}.

Conversely, eternal inflation provides a mechanism to `populate the landscape' \cite{guth3}, another being that of the sum-over-histories (no-boundary) approach \cite{haw-her}.

These two versions of a multiverse are part of a broader classification given in \cite{teg}, where many other speculative versions of the idea of a multiverse as discussed.

An open question here is in what sense all these different `universes' are really distinct to each other and how one would be able to avoid counting same ones as different. The general problem of how such a structure may evolve is also unknown,  but of course no one really knows the nature of what holds the multiverse together, or how M-theory operates in this `moduli space of supersymmetric vacua' (to quote \cite{suss}). This is one of the frontiers of theoretical mathematical cosmology.

\subsection{Topology and cosmology}\label{top}
Another quantum possibility (or rather more like a speculation at the time) was suggested in 1973 by E. Tryon \cite{try}. He imagined the universe as a quantum fluctuation of the vacuum, appearing accidentally so to speak `from time to time' and obeying Heisenberg's uncertainty principle. Thus an infinite lifetime would be associated with zero energy, but the unexplained issue with this proposal was why the universe had such a large age. But what would be the properties of a universe created from a fluctuation of the vacuum? A finitely-born universe would have all possible shapes, and the Einstein equations only relate the local properties of spacetime to the overall matter density. The issue of \emph{topology of the universe} relates to global properties, possibly observable, and in fact something like this could alter the images of observed galaxies by producing multiple, fainter  copies of their images.

D. Sokolov and V. Shvartsman in 1974 \cite{so-sh}, and R. Gott in 1980 \cite{gott1} gave lower bounds on the size of a finite universe. Also Y. Zeldovich and A. A. Starobinski in 1974 \cite{zstar} studied a flat, finite,  universe as a quantum fluctuation and concluded that it must have an approximate spherical shape to avoid a singularity. Another issue with all these studies was how to create an infinite model with global non-trivial topology. It is difficult not to take seriously any model that allows for some mild form of non-trivial global topology. Today the problem of observationally detecting some aspect of large-scale topology is an active one \cite{luminet}.

As we discuss in more detail in  Section \ref{generic}, the number of arbitrary spatial functions to describe the generality of a cosmology is 4 in vacuum for Einstein's gravity and becomes 16 for higher-order gravity.  This in turns  becomes $4(1+F)+2S$ and $16+4F+2S$ respectively,  when a number $F$ of fluids and a number $S$ of scalar fields is added, cf. \cite{ba14} (this counting assumes that dark energy is described by a cosmological constant).

It is an interesting result which we now discuss that in the case of a Bianchi (homogeneous) cosmology having compact  spatial topology - where because of the homogeneity assumption the arbitrary functions become constants -  the corresponding numbers of constants require to determine the general solution generally increase without bound. This constitutes a major difference with respect to the corresponding situation of `trivial' topologies. In the general inhomogeneous case, the situation is largely unknown.

The Einstein equations do not impose any restrictions on the spatial topological complexity of a solution, and this means that one may arbitrarily change the spatial topology of an original spacetime-solution of the field equations and still keep the resulting configuration as a solution. Topological complexity, that is the complexity of a given topology of some solution to the field equations, is something that is measured by the number of the moduli degrees of freedom, namely, the number $N_M=6g+2k-6$, where $g$ denotes the genus and $k$ the number of conical singularities of the underlying orbifold. (Although every manifold is trivially an orbifold, the converse is not true. To get an idea of what an orbifold very roughly looks like, take a properly discontinuous action of a group $\Gamma$ acting on a manifold $M$, then an orbifold looks locally like the  coordinate system chart $M\rightarrow M/\Gamma$. It is an orbit-manifold.)

Returning to our discussion of Bianchi spatial topologies, it is an intriguing fact that when compact spatial topologies are admitted to the standard Bianchi universes, the well-known  properties met in standard Bianchi cosmology change to such a degree that the situation becomes almost reversed (see \cite{fa}-\cite{ba-ko01} for more information). Some Bianchi universes (IV and VI$_h$) no longer exist, while those that contained the FRWL universes and were the generic ones with trivial topologies, now are no longer generic. For example, the VII$_h$ universe must be isotropic and hence not generic, and the Bianchi IX universe with compact topology of any complexity is nongeneric, but B-III and B-VIII now become the most general types, although they do not contain FRWL universes as special cases and get arbitrarily close to them.

The conclusion of the consideration of anisotropic universes with compact topologies is that they become heavily restricted, while isotropic ones are nongeneric. Compact open universes become necessarily  isotropic if they are assumed homogeneous, and flat spaces are generally preferred than closed or open ones.

The consideration of inhomogeneities and/or  higher dimensionality may lead to further restrictions in future cosmological theories that may not be obtained otherwise except by such topological considerations. This is because there is an intimate relation between inhomogeneity on a cosmologically large scale and topology \cite{fm1}. We recall that the degree of a manifold, $deg M$, equal to the dimension of its isometry group,  is a number that shows the generality of it, and for compact manifolds (like the ones we have been considering in this Section) $deg M=0$ if and only if there is no nonzero global Killing field. Now according to Bochner's theorem, compact manifolds having negative Ricci curvature admit no nonzero global Killing vectorfields which in turn implies that they can only be \emph{locally} homogeneous. We note that from local isotropy (say about us) one can only conclude \emph{local} homogeneity (because of constant curvature), not global one as it is usually assumed in the Bianchi class.

The broader class of locally homogeneous cosmologies becomes  therefore a very interesting area of study in mathematical cosmology (cf. \cite{fm2} and refs. therein). This subject closely ties and hugely extends the area of Bianchi topology through an analysis of hamiltonian cosmology in this context. The results are truly remarkable and lead to a completely new picture of what topological structures may arise in a global situation under the Einstein evolution. A pivotal role is played by the mathematical theory of symplectic reduction giving in the present context a reduced hamiltonian function for cosmology, such that asymptotically in the future direction of cosmological expansion the evolution is dominated by the so-called hyperbolizable components, on each one of which the conformal geometry described by a suitable metric becomes homogenous and isotropic, locally indistinguishable from a negatively curved FRWL domain, cf. \cite{fm3} for a review and \cite{mm1} for more recent work on this problem.

Understanding the global behaviour of  solutions to any set of cosmological equations (such as the Einstein equations) and their dependence upon changes of spatial topology is one of the most important problems of mathematical cosmology.

\subsection{Dynamical singularities}
As we have already indicated, the nature of cosmological singularities is a problem left out by the singularity theorems of general relativity, or by the corresponding ones in eternal inflation. The issue of the nature of singularities is indeed a very complex one, having many facets, and has recently received a further boost of activity, which we shall now briefly discuss.

There is a classification of homogeneous universes reviewed in Collins and Ellis \cite{ce79} exploiting techniques used in the singularity theorems (e.g., Raychaudhuri equation) as well as qualitative studies of the field equations themselves (for the purposes of this classification the field equations are ordinary differential equations so that standard dynamical systems techniques may be applied).  In this case, there  arise  infinite density singularities as well as infinities in certain  curvature invariants in all  FRWL and orthogonal Bianchi universes. An interesting aspect of this classification is however, that the big bang is not necessarily a point singularity, but other types are possible, such as pancakes, cigars, etc.

The situation changes completely in tilted Bianchi models (except type IX) where, apart from big bang singularities, there are now finite density ones, where there is the intriguing possibility that the fluid flow may continue past the finite density singularity, as the flow lines turn null and a Cauchy horizon is developed where the evolutions comes to an end (cf. Fig. 1, and Section 8 of \cite{ce79}). This behaviour may be very clearly seen in type V models (which contain the open FRWL universes), and the overall behaviour is very different from that occurring in the simplest isotropic models for a variety of fluids. For Bianchi type IX universes instead, we have the closed universe recollapse theorem of R. M. Wald which states that there do not exist any eternally expanding Bianchi IX universes with matter satisfying the dominant energy condition, and has non-negative average pressure \cite{wald-rec}.

There is however another type of singularity classification which is based on  the work of Y. Choquet-Bruhat and stems from theorems giving sufficient conditions for causal g-completeness, that is necessary conditions for the development of cosmological singularities (cf. \cite{cbc}, \cite{ycb} chap. 8). This leads directly to a complete dynamical classification for isotropic cosmologies into many distinct types \cite{ck1,ck2}. In fact, there are four types of this classification that are found to play an important role in current models, as they appear  commonly in dark energy universes  with phantom fluids \cite{no}.

This leads to a `zoo' of cosmological singularities developing in finite time into the future, already in the isotropic category, to say nothing about possibilities in more general homogeneous  universes. These singularities include not only the traditional ones discussed earlier (big bang and big crunch) but many other types, such as big rips, sudden singularities, or very mild soft ones, turnarounds, etc (see \cite{cot1} for a brief review). They  can be studied by using generalized power series \cite{laz1,vis1}, or the method of dominant balance \cite{bc2}, and further classified using the notion of strength of singularities. These studies have provided a clear picture of the dynamical behaviours possible, but are  limited only in the isotropic category.

\subsection{Generic universes}\label{generic}
As we have already discussed earlier, a common property of most of the cosmological solutions  met in Einstein's theory is the occurrence of infinite curvatures and densities a finite time in their past, a `cosmological singularity'. Despite lacking a rigorous definition of what a singularity is at the time, E. M. Lifshitz and I. M. Khalatnikov in 1963 (LK hereafter) were able to show that the \emph{generic singularity} present in typical solutions of the Einstein equations cannot have the usual simple, monotone, power-law character predicted by isotropic (or simple anisotropic) solutions \cite{bkl1}. Their analysis also extended   the Lifshitz 1946 instability study to the more general situation where the examined perturbations were now comparable to the horizon scale (proportional to the scale factor in a Friedmann universe).

In their  analysis, LK introduced a method of proof that was based on series expansions around some unperturbed state (vacuum, or simple radiation solutions), and then counting the arbitrary functions present in the resulting solution of the field equations considered in the synchronous reference system. If that number was equal to the maximum allowed number by the initial value problem, they concluded that the solution was a general one, otherwise the solution was a special one based on a smaller number of free initial data (compare with Section \ref{top}). For instance, the number of free initial data required for a general solution is equal to 4 for vacuum general relativity.

This function counting method is more useful than it is expected naively. There are solutions with the required number of arbitrary functions (in terms of function counting) to quantify as general solutions of the field equations of a given theory, and others that have smaller such numbers. In the latter case, the solutions possess a `transient' nature, they are unstable. In the former case, they are general solutions of the field equations, and consequently stable in the time intervals they are taken. For example, full functions counting solutions include the perturbed de Sitter space found in \cite{star1}, the sudden finite-time singularity solution in general relativity \cite{bcts}, and in Brans-Dicke theory \cite{bctr},  and the ultrastiff perfect fluid solution having $p > \rho$ near quasi-isotropic singularities \cite{hs}. On the other hand, unstable, transient solutions are more common, for example the standard quasi-isotropic solutions  in general relativity is unstable to perturbations within that theory, cf. \cite{ll}, p. 368, \cite{kl1,kl2}, and so are solutions  containing one or more fluids in that theory, \cite{kkms, kks}. For other approaches to the function counting problem, see \cite{dl}, \cite{to} and refs. therein.

The functions counted in these problems have simple physical interpretations. For example, the four free parameters in general relativity specifying a general solution in vacuum describe  two shear modes plus anisotropies in the spatial curvature. When matter is added, the extra functions are the density or pressure, and the three (non-comoving) fluid velocity components, and although these contribute to the temperature fluctuations of the CMB radiation, they can be usually ignored being too small to be detected presently. However, many of the more elaborate cosmological models discussed in this paper, e.g., eternal inflation or M-theoretic models, predict effects that are in principle described by many arbitrary functions of the sort appearing in function counting problems. For inflation, they are  also unobservable lying beyond our particle horizon, albeit isotropic.

\subsection{Dark energy}
\emph{Dark energy} representing the major component of the current acceleration of the universe has become the prime focus of attention since 1998 when two groups \cite{group1,group2} using different data sets and techniques discovered that observations showed that the speed of the expansion of distant supernovae curved upwards with relation to their distance.
This discovery has had an immense impact on cosmology ever since, and confirmed the expectations of a cosmological constant - the so-called $\Lambda$CDM (CDM here stands for `cold dark matter', cf. \cite{pee}, Section 7.2) phenomenological model - but allowed for the existence of many, more exotic stresses to describe this mysterious kind of energy.

Today models of dark energy span the whole range of possibilities, from a cosmological constant, to scalar fields - the so-called quintessence - to modified gravity theories. It is indeed an amazing feature of this hypothetical type of energy the fact that it has unified in a single quest the incredible spectrum of modified gravity theories and other exotic forms of energy to search for the missing explanation of this observed effect.

Dark energy appears as a property inherent in spacetime because it is perfectly uniform and insensitive to spacetime being empty or full of galaxies, or with respect to the direction we look, or the era in the universe history, or spatial location. The constraints from the supernovae observations, the CMB temperature fluctuations and the baryon acoustic oscillations (that is residual sound waves imprinted in the clustering patterns) all converge to the astonishing dark energy budget of about 72 per cent in the overall matter energy content of the universe.
Another component of the remaining material in the universe consists of dark matter with a percentage of about another 24 per cent.
The mysterious fact that the visible components of matter and radiation may constitute only about a tiny 4 per cent of the overall distribution, represents perhaps the most unbelievable, unexplained result in the whole history of astronomy.

We thus see that the fundamental themes required for an adequate explanation of the cosmological constant, the energy of the vacuum, and the nature of dark energy are all linked together as suggested by both theory and observations. The completely and seemingly disparate and independent fields such as scalar-tensor theory, higher-order gravity, theories with screening, Palatini cosmology, effective theories, Horndeski gravity,  higher-dimensional cosmology, Born-Infeld type theories, holographic theory, and many others, appear as one field having different facets in this sense, for various excellent reviews of this vast field, see \cite{copeland,clifton,capo,joyce,olmo1,olmo1,horn,eff,holo}.

\section{Contents and abstracts  of the Theme Issue}
Theoretical mathematical cosmology, a most majestic of fields, is not a subject that an interested person may easily enter. When one overcomes the difficulties of acquiring a proper or needed background, one will need a clear compass as to what research problems and directions are available, important, or doable.

In previous Sections, we have given some taste of the amazing variety of research in this field by discussing some of the most important models, theories and remaining problems in the field. We have focused on various cosmological models, such as, the Einstein static universe, de Sitter space, Lema\^{i}tre universe, Einstein-de Sitter universe, Tolman's oscillation model, the Friedmann family of cosmologies, the Kasner universe, the Lema\^{i}tre-Tolman inhomogeneous model, Lifshitz perturbed universes, G\"{o}del spacetime, and many other universes.

We have given short descriptions of various cosmological theories such as, the hot big bang, the steady-state theory, chaotic cosmology, inflation, the multiverse, quantum universes, Kaluza-Klein and other multi-dimensional worlds, stringy universes, braneworlds, and the landscape. These theories in turn use for their formulations physical theories that aim at describing gravitation at various stages of the evolution of the universe, from general relativity, to modified gravity, to string and M-theory, to dark energy.

However, many theoretical challenges remain. The singularity problem is one of them, but we have also included discussions of the BKL conjecture, the measure problem, the stability issue, topological problems, the classification and nature of dynamical singularities, the horizon, flatness and entropy problems, and the problem of genericity in cosmology.

Research today in mathematical cosmology is active in all these aspects of the field, while it appears that research papers in mathematical cosmology are scattered in diverse publications, together with other, often very different research, and sometimes it becomes easy to lose orientation. We therefore end this paper with the inclusion of the abstracts of all contributions to the Theme Issue `The Future of Mathematical Cosmology', in an effort to give the reader some idea as to where research in mathematical  and theoretical cosmology will be heading today and in the coming years.

Many of the  invited contributors of this Theme Issue are leaders of research in this area. We believe that their collective presence in this volume will provide a direct proof of the richness and importance of the whole field, as well as the required orientation, without further need of defence.  We hope that the scientific community will be interested to see what is included in this collection of papers, the titles and abstracts of which are reproduced in this Section (ref. numbers in these abstracts correspond to those in the specific papers), hoping that this Theme Issue may become a unique pole of attraction and a point of reference for future developments.

This Theme Issue contains many of the important results and meaningful research directions of the future, their connections to the past, as well as interconnections to other independent but important issues of the same field, that researchers may not easily have in their disposal or discover by themselves. Works appearing here have a direct relevance with the various topics touched in this paper, and we hope their diverse nature will look  somewhat more coherent  by including these abstracts here.

Interest in theoretical mathematical cosmology was recently renewed by the decision to award the 2021 Nobel Prize in Physics to R. Penrose for his work on the proof of the first singularity theorem in 1965, a result which appears as one of the central pillars of the whole field of modern mathematical cosmology as we showed in the previous Sections. This Theme Issue appearing in the Philosophical Transactions of the Royal Society A as an  edited collection of important results, problems, and future research directions in this field will, we therefore hope, influence the wider scientific community as well as policy makers to further appreciate its unique importance for the future development of this fundamental mathematical science, theoretical cosmology.

\subsection{Volume 1}
\begin{enumerate}
\item  S. Scott and P. Threlfall, Cosmological Milestones, Conformal Frameworks and Quiescent Cosmology

To understand the nature of the birth of our Universe and its eventual demise is a driving force in theoretical physics and, indeed, for humanity. A zoo of definitions has appeared in the literature to catalogue different types of cosmological milestones such as `Big Bangs', `Big Crunches', `Big Rips', `Sudden Singularities', `Bounces' and `Turnarounds’. Quiescent cosmology is the notion that the Universe commenced in a Big Bang that was highly regular and smooth, and evolved away from this initial isotropy and homogeneity due to gravitational attraction. The quiescent cosmology concept meshes well with Penrose's ideas regarding gravitational entropy and the clumping of matter, and the associated Weyl Curvature Hypothesis. Conformal frameworks, such as the Isotropic Past Singularity, have been devised to encapsulate initial and final states for the Universe which are in accordance with these programs. Since much of the research on cosmological milestones has been focussed on FRW solutions, many of which possess initial singularities which are isotropic Big Bangs, we analyse here the relationship between cosmological milestones and conformal frameworks for these solutions. We establish the general properties of FRW models which admit these conformal frameworks, including whether they satisfy various energy conditions, and are therefore physically reasonable.

\item  V. Moncrief and P. Mondal, Einstein flow with matter sources: stability and convergence

Two recent articles [1, 2] suggested an interesting dynamical mechanism within the framework of the vacuum Einstein flow (or Einstein-$\Lambda$ flow if a positive cosmological constant $\Lambda$ is included) which suggests that many closed (compact without boundary) manifolds that do not support homogeneous and isotropic metrics at all will nevertheless evolve to be asymptotically compatible with the observed approximate homogeneity and isotropy of the physical universe. These studies however did not include matter sources. Therefore the aim of the present study is to include suitable matter sources and investigate whether one is able to draw a similar conclusion.

\item  E. Ames,  F. Beyer, J. Isenberg, and T. A. Oliynyk, Stability of Asymptotic Behavior Within Polarised T2-
Symmetric Vacuum Solutions with Cosmological Constant

We prove the nonlinear stability of the asymptotic behavior of perturbations of subfamilies of Kasner
solutions in the contracting time direction within the class of polarised $T^2$-symmetric solutions of the vacuum
Einstein equations with arbitrary cosmological constant $\Lambda$. This stability result generalizes the results proven
in [3], which focus on the $\Lambda= 0$ case, and as in that article, the proof relies on an areal time foliation and
Fuchsian techniques. Even for $\Lambda= 0$, the results established here apply to a wider class of perturbations of
Kasner solutions within the family of polarised $T^2$-symmetric vacuum solutions than those considered in [3] and
[26]. Our results establish that the areal time coordinate takes all values in $(0, T_0]$ for some $T_0 > 0$, for certain families of polarised T2-symmetric solutions with cosmological constant.

\item  J.M.M. Senovilla, A critical appraisal of the singularity theorems

The 2020 Nobel prize in Physics has revived the interest in the singularity theorems and, in particular, in the Penrose theorem published in 1965. In this short paper I briefly review the main ideas behind the theorems and then proceed to an evaluation of their hypotheses and implications. I will try to dispel some common misconceptions about the theorems and their conclusions, as well as to convey some of their rarely mentioned consequences. In particular, a discussion of spacetime extensions in relation to the theorems is provided. The nature of the singularity inside black holes is also analyzed.

\item T. Padmanabhan, Lessons from the cosmological constant about the nature of gravity

Brief description: The existence of a positive cosmological constant allows us to infer several key features about the true nature of gravity. I will describe how this approach demands a paradigm shift in our description of gravity and suggests that gravity is an emergent phenomenon.

\emph{Editors' Comment: It was with great sorrow that we learnt that Thanu Padmanabhan passed away on September 17 September 2021, a few days before he was  to submit his invited contribution to this Theme Issue. We very much regret the loss of Paddy, a great cosmologist and an amazing human being. We include here the original title and abstract of his contribution, which he sent to us in an email on 18 February 2021.}

\item  K. A. Bronnikov, Some unusual wormholes in general relativity

In this short review we present some recently obtained traversable wormhole models in the framework of general relativity (GR) in four and six dimensions that somehow widen our common ideas on wormhole existence and properties. These are, first, rotating cylindrical wormholes, asymptotically flat in the radial direction and existing without exotic matter. The topological censorship theorems are not violated due to lack of asymptotic flatness in all spatial directions. Second, these are cosmological wormholes constructed on the basis of the Lemaitre-Tolman-Bondi solution. They connect two copies of a closed Friedmann world filled with dust, or two otherwise distant parts of the same Friedmann world. Third, these are wormholes obtained in six-dimensional GR, whose one entrance is located in "our" asymptotically flat world with very small extra dimensions while the other "end" belongs to a universe with large extra dimensions and therefore different physical properties. The possible observable features of such wormholes are briefly discussed.

\item  S. Cotsakis, Onset of synchronization in coupled Mixmaster oscillators

We consider the problem of asymptotic synchronization of different spatial points coupled to each other in inhomogeneous spacetime and undergoing chaotic Mixmaster oscillations towards the singularity.
We demonstrate that for couplings larger than some threshold value,  two Mixmaster spatial points $A,B$, with $A$ in the past of $B$, synchronize and thereby proceed in perfect unison towards the initial singularity.
We further show that there is a Lyapunov function for the synchronization dynamics that makes different spatial points able to synchronize exponentially fast in the past direction.
We provide an elementary proof of how an arbitrary spatial point responds to the mean field created by the oscillators, leading to their direct interaction through spontaneous synchronization.
These results ascribe a clear physical meaning of early-time synchronization  leading to a resetting effect for the two BKL maps corresponding to  two distinct oscillating spatial points, as the two maps converge to each other to  become indistinguishable at the end of synchronization. Our results imply that the universe generically organizes itself  through  simpler, synchronized, states as it approaches the initial singularity.
A discussion of further implications of early-time inhomogeneous Mixmaster synchronization is also provided.

\item  Ruth Lazkoz and Leonardo Fern\'{a}ndez-Jambrina, New futures for cosmological models

The discovery of accelerated expansion of the universe opened the
possibility of new scenarios for the doom of our spacetime, besides
aeternal expansion and a final contraction.  In this paper we review
the chances which may await our universe.  In particular, there are
new possible singular fates (sudden singularities, big rip...), but
there also other evolutions which cannot be considered as singular.
In addition to this, some of the singular fates are not strong enough
in the sense that the spacetime can be extended beyond the
singularity.  For deriving our results we make use of generalised
power expansions of the scale factor of the universe.

\item  V. Ivashchuk, On stable exponential cosmological solutions with two factor spaces in $(1+ m + 2)$-dimensional EGB model with $\Lambda$-term

A $(m+ 3)$-dimensional Einstein-Gauss-Bonnet gravitational model including the Gauss-Bonnet term and the cosmological term $\Lambda$ is considered. Exact solutions with exponential time dependence of two scale factors, governed by two Hubble-like parameters $H >0$ and $h \neq H$, corresponding to factor spaces of dimensions $m >2$ and $l = 2$, respectively, are found. Under certain restrictions on $x = h/H $, the stability of the solutions in a class of cosmological solutions with diagonal metrics is proved.
A subclass of solutions with small enough variation of the effective gravitational constant $G$ is considered and the stability of all solutions from this subclass is shown.

\item N. Mavromatos, Geometrical origins of the Universe dark sector: string-inspired
torsion and anomalies as seeds for inflation and
dark matter

In a modest attempt to present potentially new
paradigms in Cosmology, including its inflationary
epoch, and initiate discussions, I review in this article
some novel, string-inspired cosmological models,
which entail a purely geometrical origin of the Dark
sector of the Universe but also of its observed
matter-antimatter asymmetry. The models contain
gravitational (string-model independent, Kalb-Ramond
(KR)) axion fields coupled to primordial gravitational
anomalies via CP-violating interactions. The anomaly
terms are four-space-time-dimensional remnants of
the Green-Schwarz counterterms appearing in the
definition of the field strength of the spin-one
antisymmetric tensor field of the (bosonic) massless
gravitational string multiplet, which also plays the
rôle of a totally antisymmetric component of torsion.
I show how in such cosmologies the presence of
primordial gravitational waves can lead to anomaly
condensates and dynamical inflation of a "runningvacuum-
model" type, without external inflatons, but
also to leptogenesis in the radiation era due to
anomaly-induced Lorentz and CPT Violating KR
axion backgrounds. I also discuss how the torsionrelated
KR-axion could acquire a mass during the
QCD epoch, thus playing the role of (a component
of) Dark Matter. Phenomenological considerations of
the inflationary and post-inflationary (in particular,
modern) eras of the model are briefly discussed,
including its potential for alleviating the observed
tensions in the cosmological data of the current epoch.

\item  G. Horndeski, Reformulating Scalar-Tensor Field Theories as Scalar-Scalar Field Theories Using a Novel Geometry

In this paper I shall show how the notions of Finsler geometry can be used to construct a similar type of geometry using a scalar field, $f$, on the cotangent bundle of a differentiable manifold, $M$. This will enable me to use the second vertical derivatives of $f$, along with the differential of a scalar field $\phi$ on $M$, to construct a Lorentzian metric tensor on $M $that depends upon $\phi$. I refer to a field theory based upon a manifold with such a Lorentzian structure as a scalar-scalar field theory. We shall study such a theory when $f$ is chosen so that the resultant metric on $M$ has the form of a Friedman-Lemaitre-Robertson-Walker metric, with the $t$ equal constant slices being flat, and $\phi$ being a function of $t$. When second-order scalar-tensor Lagrangians are evaluated for this choice of geometry, they give rise to Lagrangians which are functions only of $\phi$ and its time derivatives. I refer to these `scalarized' Lagrangians as `Hidden Lagrangians'. However, not all Lagrangians used in scalar-scalar field theories need come from scalar-tensor Lagrangians. A particularly simple `pure' scalar-scalar Lagrangian will be investigated in detail. It will be shown that this Lagrangian can generate self-inflating universes, which can be pieced together to form multiverses with non-Hausdorff topologies in which the global time function multifurcates at $t=0$. Some of the universes in these multiverses begin explosively, and then settle down to a period of much quieter accelerated expansion, which can be followed by a collapse to their original, pre-expansion state.
\end{enumerate}

\subsection{Volume 2}
\begin{enumerate}

\item  R. Brandenberger, Limitations of an Effective Field Theory Treatment
        of Early Universe Cosmology

Assuming that superstring theory is the fundamental theory which unifies all forces of Nature at the quantum level, I argue that there are key limitations on the applicability of effective field theory techniques in describing early universe cosmology.

\item 	M. Gasperini, From pre- to post-big bang: an (almost) self-dual
cosmological history

We present a short introduction to a non-standard
cosmological scenario motivated by the duality
symmetries of string theory, in which the big bang
singularity is replaced with a "big bounce" at high but
finite curvature. The bouncing epoch is prepared by
a long (possibly infinitely extended) phase of cosmic
evolution, starting from an initial state asymptotically
approaching the string perturbative vacuum.

\item 	I. Antoniadis, S. Cotsakis, and I. Klaoudatou, Brane-world singularities and asymptotics in a five-dimensional fluid bulk

We review studies on the singularity structure and
asymptotic analysis of a 3-brane (flat or curved) embedded in a
five-dimensional bulk filled with a `perfect fluid’ with
an equation of state $p=\gamma\rho$, where $p$ is the ‘pressure’ and $\rho$ is the `density’
of the fluid, depending on the 5th space coordinate. \\
Regular solutions satisfying positive energy conditions in the bulk exist only in the cases of a flat brane
for $\gamma=-1$ or of AdS branes for $\gamma\in [-1,-1/2)$. More cases can be found by gluing two
regular brunches of solutions at the position of the brane. However, only a flat brane
for $\gamma=-1$ leads to finite Planck mass on the brane and thus localises gravity. In a more recent work, we showed that a way to rectify the previous findings and obtain
a solution for a flat brane and a range of $\gamma$, that is both free from finite-distance singularities
and compatible with the physical conditions of energy and finiteness of
four-dimensional Planck mass, is by introducing a bulk fluid component that satisfies a non-linear
equation of state of the form $p=\gamma\rho^{\lambda}$ with $\gamma<0$ and $\lambda>1$.

\item  P. Vanhove, A $S$-matrix approach to gravitational waves physics

The observation of gravitational waves emitted by binary systems has opened a new astronomical window into the Universe. We describe recent advances in the field of scattering amplitudes applied to the post-Minkowskian expansion, and the extraction of the effective two-body gravitational potential. The techniques presented here apply to any effective field theory of gravity and are not restricted to four-dimensional Einstein gravity.

\item  J. Sola, Cosmological Constant Problem and Running Vacuum in the Expanding Universe

It is well-known that quantum field theory (QFT) induces a huge value of the cosmological constant, $\Lambda$, which is outrageously inconsistent with cosmological observations. We review here some aspects of this fundamental theoretical conundrum (`the cosmological constant problem') and strongly argue in favor of the possibility that the cosmic vacuum density $\rho_{\rm vac}$ may be mildly evolving with the expansion rate $H$. Such a `running vacuum model' (RVM) proposal predicts an effective dynamical dark energy without postulating new ad hoc fields (quintessence and the like). Using the method of adiabatic renormalization within QFT in curved spacetime we find that $\rho_{\rm vac}(H)$ acquires a dynamical component ${\cal O}(H^2)$ caused by the quantum matter effects. There are also ${\cal O}(H^n)$ ($n=4,6,..$) contributions, some of which may trigger inflation in the early universe. Remarkably, the evolution of the adiabatically renormalized $\rho_{\rm vac}(H)$ is not affected by dangerous terms proportional to the quartic power of the masses ($\sim m^4$) of the fields. Traditionally, these terms have been the main source of trouble as they are responsible for the extreme fine tuning feature of the cosmological constant problem. In the context under study, however, $\rho_{\rm vac}(H)$ is currently dominated by a constant term plus the aforementioned mild dynamical component $\sim \nu H^2$ ($|\nu|\ll1$), which makes the RVM to mimic quintessence.

\item  C. W. Erickson, Rahim Leung, and K. S. Stelle, Higgs Effect Without Lunch

Reduction in effective spacetime dimensionality can occur in field-theory models more general than the widely studied dimensional reductions based on technically consistent truncations. Situations where wavefunction factors depend nontrivially on coordinates transverse to the effective lower dimension can give rise to unusual patterns of gauge symmetry breaking. Leading-order gauge modes can be left massless, but naturally occurring Stueckelberg modes can couple importantly at quartic order and higher, thus generating a “covert” pattern of gauge symmetry breaking. Such a situation is illustrated in a five-dimensional model of scalar electrodynamics in which one spatial dimension is taken to be an interval with Dirichlet/Robin boundary conditions on opposing ends. This simple model illuminates a mechanism which also has been found in gravitational braneworld scenarios.

\item P.K. Townsend, Aether, Dark Energy, and String Compactifications

The 19th century Aether died with Special Relativity but was resurrected by General
Relativity in the form of dark energy; a tensile material with tension equal to
its energy density. Such a material is provided by the D-branes of string-theory;
these can support the fields of supersymmetric particle-physics, although their en-
energy density is cancelled by orientifold singularities upon compactification. Dark
energy can still arise from supersymmetry-breaking anti-D-branes but it is probably
time-dependent. Recent results on time-dependent compactifications to an
FLRW universe with late-time accelerated expansion are reviewed.

%
%

\item  L. Tan, N. C. Tsamis and R. P. Woodard, How Inflationary Gravitons
Affect Gravitational Radiation

We include the single graviton loop contribution to
the linearized Einstein equation. Explicit results are
obtained for one loop corrections to the propagation
of gravitational radiation. Although suppressed by a
minuscule loop-counting parameter, these corrections
are enhanced by the square of the number of
inflationary e-foldings. One consequence is that
perturbation theory breaks down for a very long
epoch of primordial inflation. Another consequence
is that the one loop correction to the tensor power
spectrum might be observable, in the far future, after
the full development of 21cm cosmology.

\item  Demosthenes Kazanas, Demetrios Papadopoulos,  Dimitris Christodoulou

We present qualitative arguments in favor of an extension of the theory of the gravitational interaction beyond that resulting from  the Hilbert-Einstein action. To this end we consider a locally conformal invariant theory of gravity, discussed some thirty years ago by Mannheim and Kazanas. We discuss its exact solution of the static, spherically symmetric configurations and, based on these, we revisit some of the outstanding problems associated with gravity, high energy interactions and sketch potential resolutions within the conformal gravity framework.

\item  Naresh Dadhich, On spacetime structure and the Universe: some issues of concept and principle

In this discourse we would like to discuss some issues of concept and principle in the context of
the following three aspects. One, how $\Lambda$ arises as a constant of spacetime structure on the same
footing as the velocity of light. These are the two constants innate to spacetime without reference
to any force or dynamics whatsoever, and are interwoven in the geometry of "free" homogeneous
spacetime. Two, how does vacuum energy gravitate? Could its gravitational interaction in principle
be included in general relativity or a new theory of quantum spacetime/gravity would be required?
Finally we would like to raise the fundamental question, how does physically the Universe expand?
Since there doesn't lie anything outside, it cannot expand into, instead it has to expand of its own
- may be by creating new space out of nothing at each instant! Thus not only was the Universe
created at some instant in the past marking the beginning in the big-bang, and at its edge it is
also continuously being created at each epoch as space expands. We thus need quantum theory of
spacetime/gravity at the both ends - ultraviolet as well as infrared.
\end{enumerate}



\addcontentsline{toc}{section}{Bibliography}
\begin{thebibliography}{99}
\bibitem{ll3}L. D. Landau and E. M. Lifshitz, \emph{Quantum Mechanics}, 3rd. ed. (Pergamon Press, 1977)
\bibitem{pee}P. J. E. Peebles, \emph{Cosmology's century} (Princeton University Press, 2020)
\bibitem{emm}G. F. R. Ellis, R. Maartens, and M. A. H. MacCallum, \emph{Relativistic cosmology }(CUP, 2012)
\bibitem{cotsakis} S. Cotsakis,  Astrophys.Space Sci.Libr. 276 (2002) 5, e-Print: gr-qc/0107090 [gr-qc]; Lect.Notes Phys. 592 (2002) 59-94, e-Print: gr-qc/0201067
\bibitem{peter}P. Peter and J-P. Uzan, \emph{Primordial Cosmology} (OUP, 2009)
\bibitem{ein0}A. Einstein, S. B. preuss. Akad. Weiss. 142 (1917). English translation in The Principle of Relativity, Eds. H. A. Lorentz, A. Einstein, H. Minkowski, and H. Weyl (Dover, 1923)
\bibitem{bern} J. Bernstein and G. Feinberg (Eds.), \emph{Cosmological Constants: Papers In Modern Cosmology} (Columbia University Press, 1989)
\bibitem{north}J. D. North, \emph{The measure of the universe: a history of modern cosmology} (Dover, 1965)

\bibitem{deS0}W. De Sitter, M. N. R. A. S. 78 (1917) 3
\bibitem{rin1}W. Rindler, \emph{Relativity: Special, General, and Cosmological}, 2nd. Ed. (OUP, 2006)
\bibitem{he} S. W. Hawking and G. F. R. Ellis, \emph{The large-scale structure of space-time} (CUP, 1973)
\bibitem{gh77}G. W. Gibbons and S. W. Hawking, Phys. Rev. D15 (1977) 2738-51
\bibitem{pe}M.J. Perry, \emph{An instability of De Sitter space,} In: The very early universe. Proceedings of the Nuffield Workshop, Cambridge, 21 June to 9 July, 1982, edited by G.W. Gibbons and S.W. Hawking and S.T.C. Siklos
(CUP, 1983)


\bibitem{cha}C. Charlier, Arkiv. f. Matematik och Fysik, 16 (1922) no. 2
\bibitem{ma1}B. B. Mandelbrot, C. R. Acad. Sci. (Paris), A280 (1975) 1551
\bibitem{har}E. Harrison, \emph{Cosmology: The science of the universe,} 2nd Ed. (CUP, 2000)
\bibitem{arny}V. I. Arnol'd, \emph{Huygens and Barrow, Newton and Hooke,} (Birkh\"{a}user, 1990), Par. 5; \emph{Mathematical Methods of Classical Mechanics}, 2nd Ed. (Springer, 1989), App. 15
\bibitem{se1}F. Selety, Annalen der Physik, 68 (1922) 281
\bibitem{devac}G. de Vaucoleurs, Science, 167 (2000) 1203
\bibitem{dy}C. Dyer, M. N. R. A. S. 189 (1979) 189
\bibitem{labini1}F. S.  Labini, M. Montuori, and L. Pietronero, \emph{Scale invariance of galaxy clustering},  Phys.Rept. 293 (1998) 61-226; e-Print: astro-ph/9711073 [astro-ph]
\bibitem{labini2}F. S.  Labini, AIP Conf.Proc. 1241 (2010) 1, 981; e-Print: 0910.3833 [astro-ph.CO]
\bibitem{labini3}G. De Marzo, F. S.  Labini,  and L. Pietronero,  Astron.Astrophys. 651 (2021) A114
\bibitem{fri1}A. Friedmann, Z. Phys. 10 (1922) 377
\bibitem{fri2}A. Friedmann, Z. Phys. 21 (1924) 326
\bibitem{sli}V. M. Slipher, Pop. Astr. 23 (1915) 21

\bibitem{le1} G. Lema\^{i}tre, Ann. Soc. Sci. Brux. A47 (1927) 49
\bibitem{le1a}G. Lema\^{i}tre, M. N. R. A. S. 91 (1931) 483
\bibitem{bookofuni}J. D. Barrow, \emph{The book of universes} (Bodley Head, 2011)
\bibitem{edd1} A. S. Eddington, M. N. R. A. S. 90 (1930) 668
\bibitem{le2}G. Lema\^{i}tre, Proc. N. A. S. 20 (1934) 12

\bibitem{ein-de}A. Einstein and W. de Sitter, Proc. Nat. Acad. Sci. USA 18 (1932) 213
\bibitem{tolman}R. Tolman, Relativity, Thermodynamics and Cosmology (Dover, 1934)
\bibitem{kasner} E. Kasner, Amer. J. Math.  43  (1921) 217

\bibitem{l1}Lema\^{i}tre, G. (1933). Ann. Soc. Sci. Bruxelles A53, 51; English transl. (1997). Gen. Rel. Grav. 29, 641.
\bibitem{t1}Tolman, R. C. Proc. Nat. Acad. Sci. USA 20 (1934) 169; repinted (1997). Gen. Rel. Grav. 29, 935.
\bibitem{es1}A. Einstein and E. G. Straus, Rev. Mod. Phys. 17 (1945) 120; \emph{ibid} 18 (1946) 148
\bibitem{er}A. Einstein and N. Rosen,  Phys. Rev. 48 (1935) 73
\bibitem{cg1}R. Capovilla and J. Guven, Phys. Rev. D52 (1995) 1072
\bibitem{ca1}B. Carter,  Int.J.Theor.Phys. 40 (2001) 2099-2130;  e-Print: gr-qc/0012036 [gr-qc]
\bibitem{mann} P. D. Mannheim, \emph{Brane-localized gravity} (World Scientific, 2005

\bibitem{lif}E. Lifshitz, J. Phys. U.S.S.R. 10, 116 (1946).
\bibitem{bkl1}V. A. Belinski, I. M. Khalatnikov, and E. M. Lifshitz, Adv. Phys. 12, 185 (1963)
\bibitem{ll2}L. D. Landau and E. M. Lifshitz, \emph{The classical theory of fields}, 4th Rev. Ed. (Pergamon Press, 1975)
\bibitem{tom}For recent work, see K. Tomita, Phys. Rev. D 71, 3504 (2005)
\bibitem{rio}N. Bartolo, S. Matarrese, and A. Riotto, J. Cosm.  Astropart. Phys. 0606, 024 (2006); e-Print:
astro-ph/0604416 [astro-ph]
\bibitem{weinberg2}S. W. Weinberg, \emph{Cosmology} (OUP, 2007)


\bibitem{ga1}G. Gamow, \emph{The evolution of the universe,} Nature 162 (30 Nov. 1948) 680
\bibitem{ga2}G. Gamow, \emph{The creation of the universe} (Viking Press, New York, 1952)
\bibitem{ah1}R. A. Alpher and R. C. Herman, \emph{Early work on the `big-bang' cosmology and the microwave background radiation,} In, Modern Cosmology in Retrospect, B. Bertotti, \emph{et al} (Eds.) (CUP, 1990)
    \bibitem{wein3}S. W. Weiberg, \emph{The first three minutes: A modern view of the origin of the universe,} 2nd Ed. (Basic Books, 1993)


\bibitem{bg}H. Bondi and T. Gold, Mon.Not.Roy.Astr.Soc. 108 (1948) 252
\bibitem{hoyle}F. Hoyle, Mon.Not.Roy.Astr.Soc. 108 (1948) 342
\bibitem{nar1}J. V. Narlikar, \emph{Introduction to cosmology} (CUP, 1993)
\bibitem{mac1}W. H.  McCrea, Proc. Roy. Soc. 296 (1951) 562
\bibitem{mac2} W. H.  McCrea, Mon.Not.Roy.Astr.Soc. 128 (1964) 335
\bibitem{pen1}R. Penrose, \emph{Cycles of time} (Bodley Head, 2010)

\bibitem{goe}K. G\"{o}del, Rev. Mod. Phys.  21 (1949) 447
\bibitem{tip}F. J. Tipler, C. J. S. Clarke, and G. F. R. Ellis, \emph{Singularities and horizons: a review article,} In, General Relativity and Gravitation: One Humdred Years after the Birth of Albert Einstein, A. Held (Ed.), volume 2 (Pleanum, 1980), p. 97
    \bibitem{ear}J. Earman, \emph{Bangs, Crunches, Whimpers, and Shrieks}(OUP, Oxford, 1995).
\bibitem{re}M.J. Reboucas and J. Tiomno, Phys. Rev. D28 (1983) 1251
\bibitem{ac}] A.J. Accioly, Nuovo Cimento B100 (1987) 703
\bibitem{bada}J. D. Barrow and M. P. Dabrowski, Phys.Rev.D 58 (1998) 103502
\bibitem{hor}E. K. Boyda, S. Ganguli, P. Horava, U. Varadarajan, Phys.Rev.D 67 (2003) 106003


\bibitem{taub}A.H. Taub, 1951, Ann. Math. 53, 472
\bibitem{bianchi}L. Bianchi, 1897, Mem. Soc. It. della. Sc. (Dei. XL) (3), 11, 267.
\bibitem{kra}A. Krasinski, \emph{Inhomogeneous cosmological models} (CUP, 1997)
\bibitem{uvwe} C. Uggla, H. van Elst, J. Wainwright and G.F.R. Ellis, Phys.Rev.D 68 (2003) 103502; arXiv: gr-qc/0304002




\bibitem{di1}P. A. M. Dirac, Nature 139 (1937) 323
\bibitem{bd1}C. Brans and R. H. Dicke, Phys. Rev. 124, 925 (1961)
\bibitem{di2}R. H. Dicke, \emph{The theoretical significance of experimental relativity} (Gordon and Breach, 1968)
\bibitem{nari1}H. Nariai, Prog. Theo. Phys. 42, 544 (1969)
\bibitem{rf1}V. A. Ruban and A. M. Filkenstein, Gen. Rel. Grav. 6 (1975) 601
\bibitem{bkl4}Belinskii, V. A. and Khalatnikov, I. M. (1973) Sov. Phys. JETP, 36, 591–7.
\bibitem{r76}V. A. Ruban, Astroph. Space Sci.  46 (1977) L23
\bibitem{ke1}S. J. Kolitch and D. M. Eardley, Ann. Phys.  241 (1995) 128
\bibitem{brans1}C. H. Brans,\emph{ Gravity and the tenacious scalar field,} In: Symposium to honor Engelbert Schucking; e-Print: gr-qc/9705069 [gr-qc]
\bibitem{brans2}C. H. Brans, \emph{The Roots of scalar-tensor theory: An Approximate history,} In the:  Santa Clara 2004: 1st International Workshop on Gravitation and Cosmology; e-Print: gr-qc/0506063 [gr-qc]
\bibitem{di3}R. H. Dicke, Phys. Rev. 125, 2163 (1962)
\bibitem{ba1}J. D. Barrow, Astrophys.Space Sci. 283 (2003) 645-660; e-Print: gr-qc/0209080 [gr-qc]
\bibitem{ba2}J. D. Barrow,  \emph{The Constants of Nature: from alpha to omega} (Jonathan Cape, 2002)
\bibitem{dn1}T. Damour and K. Nordtvedt,  Phys.Rev.D 48 (1993) 3436-3450
\bibitem{mn1}J. P. Mimoso and A. M. Nunes, Phys.Lett.A 248 (1998) 325-331
\bibitem{gasp}M. Gasperini, \emph{Elements of String Cosmology} (CUP, 2007)
\bibitem{cartan}E.Cartan,  Annales scientifiques de l’E.N.S. 3e s\'{e}rie, tome 41 (1924) 1
\bibitem{kibble}T.W.B.Kibble,  J. Math. Phys., 2 (1961) pp. 212-221

\bibitem{sciama}	D.W.Sciama,  Rev. Mod. Phys., 36 (1961) pp. 463-469
\bibitem{Kopczynski}W.Kopczyński,  Phys. Lett. A39 (1972) pp. 219–220
\bibitem{Trautman}A.Trautman,  Nature, 242 (1973) 7
\bibitem{hehl1}F.W.Hehl, P. von der Heyde, G.D.Kerlick,  Rev. Mod. Phys., 48 (1976) pp. 393-416
\bibitem{hehl2}F. W. Hehl, J. D.  McCrea, E. W. Mielke, Y. Ne'eman, Phys.Rept. 258 (1995) 1-171; e-Print: gr-qc/9402012 [gr-qc]
\bibitem{sab2}V. de Sabbata and M. Gasperini,  Lettere al Nuovo Cimento, Vol. 28, (1980) pp. 229-233
\bibitem{sab1}V. de Sabbata and M. Gasperini, Lettere al Nuovo Cimento, Vol. 28 (1980) pp. 234-236
\bibitem{sab3}V. de Sabbata and M. Gasperini,  Lettere al Nuovo Cimento, Vol. 27 (1980) pp. 289-292
\bibitem{yef1}A. P. Yefremov,  Acta Phys. Pol., 3 (1981) pp.185-188
\bibitem{sari1}Yi-Fu  Cai, S. Capozziello, M.  De Laurentis,  and  E.  N. Saridakis, Rept.Prog.Phys. 79 (2016) 10, 106901; e-Print: 1511.07586 [gr-qc]
\bibitem{to1}G. R. Bengochea and R. Ferraro, Phys.Rev.D 79 (2009) 124019; e-Print: 0812.1205 [astro-ph]
\bibitem{to2}E. V. Linder, Phys.Rev.D 81 (2010) 127301, Phys.Rev.D 82 (2010) 109902 (erratum); e-Print: 1005.3039 [astro-ph.CO]
\bibitem{cr1}L. Combi, G. E. Romero, Annalen Phys. 530 (2018) 1, 1700175; e-Print: 1708.04569 [gr-qc]
\bibitem{yang1}R.-J. Yang,		Europhys.Lett.93:60001,2011; arXiv:1010.1376 [gr-qc]
\bibitem{mis4}C. W. Misner, J. Math. Phys. 7 (1963) 924
\bibitem{pen}R. Penrose, Phys. Rev. Lett. 14 (1965) 57
\bibitem{haw}S. W. Hawking, Proc. R. Soc. Lond. A 294 (1966) 511; \emph{ibid} 295 (1966) 490; \emph{ibid} 300 (1967) 187
\bibitem{hp}S. W. Hawking and R. Penrose, Proc. R. Soc. Lond. A 314 (1970) 529
\bibitem{ge1}R. Geroch, Ann. Phys. 48 (1968) 526
\bibitem{ge2}R. Geroch, J. Math. Phys. 11 (1970) 437
\bibitem{ge3}R. Geroch, Gen.Rel. Grav. 2 (1971) 61
\bibitem{ellis}G. F. R. Ellis, Gen.Rel. Grav. 2 (1971) 7
\bibitem{penlect}R. Penrose, \emph{Techniques of differential topology in relativity} (SIAM, 1972)
\bibitem{khalatnikov}I.M. Khalatnikov, A.Yu. Kamenshchik, Phys.Usp. 51 (2008) 609-616; e-Print: 0803.2684 [gr-qc]
\bibitem{belinski}V. A. Belinski, Int.J.Mod.Phys.D 23 (2014) 1430016; e-Print: 1404.3864 [gr-qc]



\bibitem{bkl2}V. A. Belinski, I. M. Khalatnikov, and E. M. Lifshitz, Adv. Phys. 19, 525 (1970)
\bibitem{misner}C. W. Misner, Phys. Rev. Lett. 22, 1071 (1969)
\bibitem{bkl3}V. A. Belinski, I. M. Khalatnikov, and E. M. Lifshitz,  Adv. Phys. 31, 639 (1982)
\bibitem{berger1}B. K. Berger and V. Moncrief,  Phys. Rev. D 48,  4676 (1993)
\bibitem{berger2}B. K. Berger and V. Moncrief, Phys. Rev. D 57, 7235 (1998)
\bibitem{berge3}B. K. Berger, V. Moncrief,  Phys. Rev. D 58, 064023 (1998)
\bibitem{berger4}B. K. Berger, D. Garfinkle, J. Isenberg, V. Moncrief, and M. Weaver, T Mod. Phys. Lett. A 13, 1565 (1998)
\bibitem{berger5}B. K. Berger, J. Isenberg,  M. Weaver,  Phys. Rev. D 64, 084006 (2001)



\bibitem{mis1}C. W. Misner, Astrophys.J. 151 (1968) 431-457
\bibitem{mis2}C. W. Misner,  Phys.Rev.Lett. 22 (1969) 1071-1074
\bibitem{adm}R. Arnowitt, S. Deser, and C. W. Misner, \emph{The dynamics of general relativity,} In, GravitationA An Introduction to Current Research, ed. L. Witten (Wiley, New York, 1962), p. 227; reprinted in GEn Rel. Grav.  40 (2008) 1997
\bibitem{mis3}C. W. Misner,  \emph{The Mixmaster cosmological metrics, }In D. Hobill Ed., Deterministic Chaos in General Relativity (Plenum Press, 1994); e-Print: gr-qc/9405068 [gr-qc]
    \bibitem{ba-ma}J. D. Barrow and R. A. Matzner, M. N. R. A. S. 181 (1977) 719
\bibitem{chitre}D. M. Chitre, \emph{Investigation of vanishing of a horizon for Bianchi type IX (the Mixmaster universe), } PhD Thesis, Technical Report No. 72-125, Univ. of Maryland (Faculty Publications Collection, 1972)
\bibitem{ba82}J. D. Barrow, Phys. Reports 85, 1 (1982)
\bibitem{ba-co89}J. D. Barrow and S. Cotsakis, Phys. Lett. B232  (1989) 172
\bibitem{co-dem93}S. Cotsakis, J. Demaret, Y. de Rop and L. Querella, Phys.
Rev. D48 (1993) 4595
\bibitem{ho91}D. Hobill, D. Bernstein, M. Welge, and D. Simkins, lass. Quant. grav. 8 (1991) 1155
\bibitem{be94}B. K. Berger, \emph{The BKL discrete evolution as an approximation to Mixmaster dynamics,} In D. Hobill Ed., Deterministic Chaos in General Relativity (Plenum Press, 1994)
\bibitem{dema}J. Demaret, Y. De Rop, Phys.Lett. B299 (1993) 223
\bibitem{bogo1}O. I. Bogoyavlenski, S. P. Novikov, Sov. Phys. JETP. 37 (1973) 747
\bibitem{we} J. Wainwright and G. F. R. Ellis, \emph{Dynamical systems in cosmology} (CUP. 1993)
\bibitem{ring}H. Ringstrom, Class.Quant.Grav. 17 (2000) 713-731; gr-qc/9911115
\bibitem{ba78}J. D. Barrow, Nature 272 (1978) 211
\bibitem{co-ha}C. B. Collins, S. W. Hawking, Astrophys.J. 180 (1973) 317-334
\bibitem{goode}S.W. Goode, J. Wainwright, Class.Quant.Grav. 2 (1985) 99-115
\bibitem{scott1}P. A. Hohn, S. M. Scott, Class.Quant.Grav. 26 (2009) 035019


\bibitem{christo}D. Christodoulou, Class. Quant. Grav.   Class. Quantum Grav. 16 (1999) A23
\bibitem{pen-road}R. Penrose, \emph{The Road to Reality }(Bodley Head, 2004)
\bibitem{witten}E. Witten Adv.Theor.Math.Phys. 2 (1998) 253-291; e-Print: hep-th/9802150 [hep-th]
\bibitem{bati88}J. D. Barrow and F. J. Tipler, Nature 331 (1988) 31
\bibitem{bafinite}J. D. Barrow,  Phys.Rev.D 101 (2020) 2, 023527; e-Print: 1912.12926 [gr-qc]
\bibitem{le-ste}J-L. Lehners and  K.S. Stelle, Phys. Rev. D 100 (2019) 8, 083540; e-Print: 1909.01169 [hep-th]
\bibitem{coleman0}S. Coleman, \emph{Aspects of symmetry} (CUP, 1985), p. 344-5
\bibitem{star0}A. A.  Starobinski, Phys. Lett. B91 (1980) 99
\bibitem{kaz}D. Kazanas, Ap. J. lett. 241 (1980) 59
\bibitem{guth1} A. H. Guth, Phys. Rev. D23 (1981) 347
\bibitem{stein0}P. J. Steinhardt,  \emph{Natural inflation,} in The Very Early Universe, Proceedings of the Nuffield Workshop, Cambridge, 21 June –9 July, 1982, eds: Gibbons, G W, Hawking, S W and Siklos, S T C (Cambridge: Cambridge University Press, 1983), pp. 251–66.
\bibitem{weinberg1}S. W. Weinberg, \emph{Gravitation and cosmology} (Addison-Wesley, 1972)
\bibitem{bran0}R. H. Brandenberger, Rev. Mod. Phys. 57 (1985) 1
\bibitem{kolb}E. W. Kolb and M. S. Turner, \emph{The Early Universe} (Addison-Wesley, 1990)

\bibitem{bicep}BICEP/Keck Collaboration, P.A.R. Ade, et al, Phys.Rev.Lett. 127 (2021) 15, 151301; e-Print: 2110.00483 [astro-ph.CO]

\bibitem{vil0}A. Vilenkin,Phys.Rev. D27 (1983) 2848
\bibitem{linde1}A. D. Linde, Phys. Lett. B 175 (1986) 395
\bibitem{linde95}A. D. Linde, \emph{Quantum cosmology and the structure of inflationary universe,} in: PASCOS / HOPKINS 1995 (Joint Meeting of the International Symposium on Particles, Strings and Cosmology and the 19th Johns Hopkins Workshop on Current Problems in Particle Theory), 349-394; e-Print: gr-qc/9508019
    \bibitem{vil3}A. Vilenkin, \emph{Eternal inflation and chaotic terminology,} e-Print: gr-qc/0409055 [gr-qc]
\bibitem{linde02}L. Kofman, A. D. Linde, V. F. Mukhanov, JHEP 10 (2002) 057; e-Print: hep-th/0206088 [hep-th]
\bibitem{wald-hollands}S. Hollands and R. M. Wald, Gen.Rel.Grav. 34 (2002) 2043-2055 • e-Print: gr-qc/0205058 [gr-qc]
\bibitem{linde2}A. D. Linde, \emph{Particle physics and inflationary cosmology} (Harwood Academic Publishers, 1990)
\bibitem{guth2}A. Borde, A. H. Guth, and A. Vilenkin, Phys. Rev. Lett. 90 (2003) 151301
\bibitem{ag}A. Aguirre and S. Gratton, 	Phys.Rev. D67 (2003) 083515; 	arXiv:gr-qc/0301042


\bibitem{DeWitt}B. S. DeWitt, Phys. Rev. 160 (1967) 1113
\bibitem{wheeler}J. A. Wheeler, in Batelle Rencontres, eds., C. DeWitrt and J. A. Wheeler (Benjamin, New York, 1968)
\bibitem{hh}J. B. Hartle and S. W. Hawking, 	Phys.Rev. D28 (1983) 2960
\bibitem{vil}A. Vilenkin, 	Phys.Rev. D33 (1985) 3560
\bibitem{vil2}A. Vilenkin, 	Phys.Rev. D37 (1987) 888
\bibitem{gib}G. W. Gibbons and L. P. Grishchuk, Nucl. Phys. B313 (1989) 736
\bibitem{hall}J. J. Halliwell, \emph{A bibliography of papers on quantum cosmology,} NSF-ITP-88-132
\bibitem{ash1}A. Ashtekar, \emph{A short review of loop quantum gravity,} Rept.Prog.Phys. 84 (2021) 4, 042001; e-Print: 2104.04394 [gr-qc]
\bibitem{bojo}M. Bojowald, \emph{Loop quantum cosmology,} Living Rev.Rel. 11 (2008) 4; e-Print: gr-qc/0601085 [gr-qc]
\bibitem{bgzk}V. A. Belinski, L. P. Grishchuk, Ya. B. Zel'dovich, and  I. M. Khalatnikov, Sov. Phys. JETP 62 (1985) 195
\bibitem{hall2} J. J. Halliwell, Phys. Lett. B185 (1987) 341
\bibitem{h1}] M. Henneaux,  Nuovo Cim. Lett. 38 (1983) 603.
\bibitem{ghs}G. W. Gibbons, S. W. Hawking and J. M. Stewart,  Nucl. Phys. B 281 (1987) 736.
\bibitem{hpage}S. W. Hawking and D. N. Page,  Nucl. Phys. B 298 (1988) 789.
\bibitem{p1}] D. N. Page,  Phys. Rev. D 36 (1987) 1607.
\bibitem{cp}] P. Chmielowski and D. N. Page,  Phys. Rev. D 38 (1988) 2392.
\bibitem{gt}G. W. Gibbons and N. Turok,  Phys. Rev. D 77 (2008) 063516, hep-th/0609095.
\bibitem{p2}D. N. Page,  JCAP 1106 (2011) 038, arXiv:1103.3699 [hep-th].
\bibitem{schi}J. S. Schiffrin and R. M. Wald,  hys.Rev.D 86 (2012) 023521; e-Print: 1202.1818 [gr-qc]
\bibitem{lin-noor}A. Linde, M. Noorbala, JCAP 09 (2010) 008; e-Print: 1006.2170 [hep-th]


\bibitem{haw-worm}S. W. Hawking, Phys.Rev.D 37 (1988) 904-910
\bibitem{coleman1}S.  Coleman, Nucl.Phys.B 310 (1988) 643-668
\bibitem{fs}W. Fischler and L. Susskind, Phys. Lett. B217 (1989)48
\bibitem{baby-proc}S. Coleman et al, (eds.), Quantum cosmology and baby universes (World Scientific, 1991)
\bibitem{mt1}M.S. Morris and K.S. Thorne, Am. J. Phys. 56, 395 (1988)
\bibitem{mt2}M.S. Morris, K.S. Thorne and U. Yurtsever, Phys. Rev. Lett. 61, 1449 (1988)
\bibitem{bro1} K. A. Bronnikov, V. N. Melnikov, and H. Dehnen, Phys.Rev.D 68 (2003) 024025; e-Print: gr-qc/0304068 [gr-qc]
\bibitem{bro2}K.A. Bronnikov, M.V. Skvortsova, A.A. Starobinsky, Grav.Cosmol. 16 (2010) 216-222; e-Print: 1005.3262 [gr-qc]
\bibitem{bro3}K. A. Bronnikov, J. P.S. Lemos, Phys.Rev.D 79 (2009) 104019; e-Print: 0902.2360 [gr-qc]
\bibitem{visser}M. Visser, \emph{Lorentzian wormholes — from Einstein to Hawking}, (AIP Press, New York, 1995)
\bibitem{cvisser}C. Molina-Paris and M Visser, Phys.Lett.B 455 (1999) 90-95; e-Print: gr-qc/9810023 [gr-qc]



\bibitem{freu}P. G. O. Freund,  Nucl. Phys. B209, 146, 1982.
\bibitem{abb} R. B. Abbott, S. M. Barr, and S. D. Ellis,  Phys. Rev., D30:720, 1984.
\bibitem{ok}Y. Okada, Nucl. Phys. B264, 197, 1982.

\bibitem{wesson}J. M. Overduin and P. S. Wesson, \emph{Kaluza-Klein Gravity}, Phys.Rept. 283, 303-380, 1997
\bibitem{furu}T. Furusawa and A. Hosoya, Prog. Theo. Phys. 73 (1985) 467
\bibitem{ba-stein}J. D. Barrow and J. A. Stein Schabes, Phys. Rev. D32 (1985) 1597
\bibitem{dem1}J. Demaret, M. Henneaux and P. Spindel, Phys. Lett. B164, 27 (1985)
\bibitem{dem2}] J. Demaret, J. L. Hanquin, M. Henneaux, P. Spindel and A. Taormina, Phys. Lett. B 175, 129 (1986)
\bibitem{dem3}J. Demaret, Y. de Rop and M. Henneaux,Phys. Lett. B211, 37 (1988)
\bibitem{hosoya}A. Hosoya, L. G. Jensen, J. A. Stein Schabes, Nucl. Phys. B283 657 (1987)





\bibitem{gasp-ven}M. Gasperini and G Veneziano, Astropart. Phys. 1 (1993) 317
\bibitem{coley1}A. P. Billyard, A. A. Coley and J. E. Lidsey,  	J.Math.Phys. 41 (2000) 6277-6283
\bibitem{ba-kun1}J. D. Barrow and K. E. Kunze, 	Phys.Rev. D55 (1997) 623-629
\bibitem{ba-kun2}J. D. Barrow and K. E. Kunze, Phys.Rev. D56 (1997) 741-752
\bibitem{wands}J. E. Lidsey, D. Wands, and E. J. Copeland, \emph{Superstring cosmology,} Phys.Rept. 337 (2000) 343-492; e-Print: hep-th/9909061 [hep-th]

\bibitem{lancz}C. Lanczos, Rev. Mod. Phys. 19 (1957) 337; \emph{ibid} 34 (1962) 379
\bibitem{buchdahl}H. A. Buchdahl, M. N. R. A. S. 150 (1970) 1
\bibitem{gu1}V. Ts. Gurovich, Sov. Phys. Doklady 15 (1971) 1105
\bibitem{nariai}H. Nariai, Prog. Theor. Phys. 49 (1973) 165
\bibitem{kerner}R. Kerner, Gen. Rel. Grav 14 (1982) 453
\bibitem{ba-ot}J. D. Barrow and A. C. Ottewill, J. Phys. A 16 (1983) 2757
\bibitem{ba-co1}J. D. Barrow and S. Cotsakis, Phys. Lett. B 214 (1988) 515
\bibitem{kaz-mann}P. D. Mannheim and D.Kazanas, Astrophys.J. 342 (1989) 635
\bibitem{cmq}S. Cotsakis, J. Miritzis, and L. Querella,  J.Math.Phys. 40 (1999) 3063-3071; e-Print: gr-qc/9712025 [gr-qc]
\bibitem{ba-coTR}J.D. Barrow, S. Cotsakis, Eur. Phys. J. C80 (2020) 839; arXiv:1907.02928



\bibitem{harrison67}E.R. Harrison, Rev. Mod. Phys. 39, 862 (1967).
\bibitem{gibbons87}G.W. Gibbons, Nucl. Phys. B292, 784 (1987); ibid., 310, 636 (1988).
\bibitem{ellis-maartens}G.F.R. Ellis, R. Maartens, Class.Quant.Grav. 21 (2004) 223-232; e-Print: gr-qc/0211082 [gr-qc]
\bibitem{tsagas}J. D. Barrow, G.F.R. Ellis, R. Maartens, C. G. Tsagas,  Class.Quant.Grav. 20 (2003) L155-L164; e-Print: gr-qc/0302094 [gr-qc]
\bibitem{barrow-yamamoto}J. D. Barrow and K. Yamamoto, Phys.Rev.D 85 (2012) 083505; e-Print: 1108.3962 [gr-qc]
\bibitem{barrow20action}J. D. Barrow, Phys.Rev.D 101 (2020) 2, 023527; e-Print: 1912.12926 [gr-qc]
\bibitem{ew}J. Wainwright and G. F. R. Ellis, \emph{Dynamical systems in cosmology} (CUP, 1993)
\bibitem{ba-so}J. D. Barrow and D. H. Sonoda, Phys. Rep. 139 (1985) 1
\bibitem{wa83}R. M. Wald, Phys. Rev. D28 (1983) 2118
\bibitem{cf1}S. Cotsakis and G. Flessas, Phys.Lett.B 319 (1993) 69-73
\bibitem{cm1}S. Cotsakis and J. Miritzis, Class.Quant.Grav. 15 (1998) 2795-2801; e-Print: gr-qc/9712026 [gr-qc]
\bibitem{ym}J. Yokoyama and K. Maeda, Phys. Rev. D41 (1990) 1047
\bibitem{wh}J. Wainwright and G.F.R. Ellis, Class. Quant. Grav. 6 (1989) 1409
\bibitem{bogo}O. I. Bogoyavlenski, \emph{Methods in the qualitative theory of dynamical systems in astrophysics and gas dynamics} (Springer, 1985)


\bibitem{bra1}S. Alexander, R. H. Brandenberger, D.A. Easson, Phys.Rev.D 62 (2000) 103509 ;e-Print: hep-th/0005212
\bibitem{ba-da1}J. D. Barrow and M. P. Dabrowski, Phys.Rev. D57 (1998)
\bibitem{dam1} T. Damour and M Henneaux, Phys.Rev.Lett. 85 (2000) 920-923
\bibitem{dam2} T. Damour and S. de Buyl, Phys.Rev. D77 (2008) 043520
\bibitem{dam3} T. Damour and Nicolai,  Int.J.Mod.Phys.D 17 (2008) 525-531; e-Print: 0705.2643 [hep-th]
\bibitem{wal1}A. Lucas, B. A. Ovrut and D. Waldram, Nucl.Phys. B495 (1997) 365-399
\bibitem{wal2}A. Lucas, B. A. Ovrut and D. Waldram, The cosmology and M-Theory of type II superstrings;	arXiv:hep-th/9802041
\bibitem{coley2}A. P. Billyard, A. A. Coley, J. E. Lidsey, and U. S. Nilsson, Phys.Rev. D61 (2000) 043504; arXiv:hep-th/9908102
\bibitem{town1}P. Townsend, Cosmic acceleration and M theory, e-Print: hep-th/0308149 [hep-th]
\bibitem{and1}L. Andersson and J. Mark Heinzle, Adv.Theor.Math.Phys. 11 (2007) 3, 371-398; e-Print: hep-th/0602102 [hep-th]
\bibitem{town2} P. Townsend and M. N. R. Wohlfarth, Class.Quant.Grav. 21 (2004) 5375; e-Print: hep-th/0404241
\bibitem{mar}D. Marolf, \emph{String/M-branes for relativists, }	arXiv:gr-qc/9908045


\bibitem{ant1}I. Antoniadis, Phys. Lett. B246 (1990) 377
\bibitem{ant2}I. Antoniadis, J. Phys. Conf. Ser. 33 (2006) 170.
\bibitem{randall}L. Randall and R. Sundrum,  Phys.Rev.Lett. 83 (1999) 4690-4693; e-Print: hep-th/9906064 [hep-th]
\bibitem{maartens}R. Maartens, \emph{Cosmological dynamics on the brane,} Living Rev.Rel. 7 (2004) 7; e-Print: gr-qc/0312059 [gr-qc]
\bibitem{langlois}D. Langlois, \emph{Brane cosmology: An introduction,} Prog.Theor.Phys.Suppl. 148 (2003) 181-212; e-Print: hep-th/0209261 [hep-th]
\bibitem{stein}J. Khoury, B. A. Ovrut, P. J. Steinhardt, N. Turok, Phys.Rev.D 64 (2001) 123522; e-Print: hep-th/0103239 [hep-th]
\bibitem{lehners}J.L. Lehners, \emph{Ekpyrotic and cyclic cosmology,} Phys.Rept. 465 (2008) 223-263 • e-Print: 0806.1245 [astro-ph]
\bibitem{bran2}R. H. Brandenberger and P. Patrick, \emph{Bouncing Cosmologies: Progress and Problems,} Found.Phys. 47 (2017) 6, 797-850; e-Print: 1603.05834 [hep-th]
\bibitem{ant-co-kl}I. Antoniadis, S. Cotsakis, I. Klaoudatou, Eur. Phys. J. C 81, 771 (2021); arXiv: 2106.15669

\bibitem{ant-co}I. Antoniadis, S. Cotsakis, Eur. Phys. J. C75 (2015) 1-12, arXiv:1409.2220.


\bibitem{suss}L. Susskind, \emph{The Anthropic landscape of string theory}, e-Print: hep-th/0302219 [hep-th]
\bibitem{haw-her}S.W. Hawking, T. Hertog, Phys.Rev.D 73 (2006) 123527, e-Print: hep-th/0602091 [hep-th]; see also J. B. Hartle, S.W. Hawking, T. Hertog, Phys.Rev.Lett. 100 (2008) 201301, e-Print: 0711.4630 [hep-th]
\bibitem{guth3}A. H. Guth, J.Phys.A 40 (2007) 6811-6826; e-Print: hep-th/0702178 [hep-th]
\bibitem{teg}M Tegmark, \emph{Parallel universes,} e-Print: astro-ph/0302131 [astro-ph]

\bibitem{try}E. Tryon, Nature 396 (1973) 246
\bibitem{so-sh}D. D. Sokolov and V. F. Shvartsman, Sov. Phys. JETP 39 (1974) 196
\bibitem{gott1}J. R. Gott, M.N.R.A.S. 193 (1980) 153
\bibitem{zstar}Ya. B. Zeldovich and A. A. Starobinski, Sov Astron. Lett. 10 (1984) 135
\bibitem{luminet}J, -P. Luminet, \emph{The wraparound universe,} (Wellesley, USA: A.K. Peters 2008)
\bibitem{ba14}J. D. Barrow, Phys.Rev.D 89 (2014) 6, 064022, e-Print: 1401.4344 [gr-qc]
\bibitem{fa}H.V. Fagundes, 1985, Phys. Rev. Lett. 54, 1200
\bibitem{ash}A. Ashtekar and J. Samuel, 1991, Class. Quantum Grav. 8, 2191
\bibitem{koi1}T. Koike, M. Tanimoto and A. Hosoya, 1993, J. Math. Phys. 35, 4855
\bibitem{koi2}M. Tanimoto, T. Koike and A. Hosoya, 1997, J. Math. Phys. 38, 350; \emph{ibid} 6557
\bibitem{ko1}H. Kodama, 1998, Prog. Theor. Phys. 99, 173-236;
\bibitem{ba-ko2}J. D. Barrow and H. Kodama,Class.Quant.Grav. 18 (2001) 1753-1766; e-Print: gr-qc/0012075 [gr-qc]
\bibitem{ba-ko01}J. D. Barrow and H. Kodama, Int.J.Mod.Phys.D 10 (2001) 785-790; e-Print: gr-qc/0105049 [gr-qc]
\bibitem{fm1} A. E. Fischer and V. Moncrief, In: \emph{ Global Structure and Evolution in General
Relativity,} S. Cotsakis and G. W. Gibbons, (eds.) Springer LNP, \textbf{460}
(1996), pp. 111-173
\bibitem{fm2} A. E. Fischer and V. Moncrief, \emph{The Reduced Hamiltonian of General Relativity and the $sigma$-constant of Conformal Geometry}, Proceedings of the 2nd Samos Meeting on Cosmology, Geometry and Relativity, Mathematical and Quantum Aspects of Relativity and Cosmology, Lecture Notes in Physics, vol. 537 S. Cotsakis and G. Gibbons (editors), (Springer-Verlag, Berlin, 2000) 70–101.
\bibitem{fm3} A. E. Fischer and V. Moncrief, \emph{Hamiltonian Reduction of Einstein's Equations,} Encyclopedia of Mathematical Physics (Academic Press, 2006), pp. 607-623
    \bibitem{mm1}V. Moncrief and P. Mondal, Pure Appl.Math.Quart. 15 (2019) 3, 921-966; e-Print: 1903.00323 [gr-qc]

\bibitem{ce79}C. B. Collins and G. F. R. Ellis, \emph{Singularities in Bianchi cosmologies,} Phys. Rept. 56 (1979) 65-105
    \bibitem{wald-rec}R. M. Wald, Phys. Rev. D41 (1990) 2444
\bibitem{cbc}Y. Choquet-Bruhat and S. Cotsakis, J.Geom.Phys. 43 (2002) 345-350 • e-Print: gr-qc/0201057 [gr-qc]
\bibitem{ycb}Y. Choquet-Bruhat, \emph{General relativity and the Einstein equations }(OUP, 2009)
\bibitem{ck1}S. Cotsakis and I. Klaoudatou, J.Geom.Phys. 55 (2005) 306-315 • e-Print: gr-qc/0409022 [gr-qc]
\bibitem{ck2}S. Cotsakis and I. Klaoudatou, J.Geom.Phys. 57 (2007) 1303-1312 • e-Print: gr-qc/0604029 [gr-qc]
\bibitem{no}S. Nojiri, S. D. Odintsov, S. Tsujikawa, Phys.Rev.D 71 (2005) 063004; e-Print: hep-th/0501025 [hep-th]
\bibitem{cot1}S. Cotsakis, \emph{Structure of infinity in cosmology,} Int.J.Mod.Phys.D 23 (2013) 1330003; e-Print: 1212.6737 [gr-qc]
\bibitem{laz1}L. Fernandez-Jambrina, R. Lazkoz, Phys.Rev.D 74 (2006) 064030; e-Print: gr-qc/0607073 [gr-qc]
\bibitem{vis1}C. Cattoen, M. Visser, Class.Quant.Grav. 22 (2005) 4913-4930; e-Print: gr-qc/0508045 [gr-qc]
\bibitem{bc2}S. Cotsakis and J. D. Barrow, \emph{The Dominant Balance at Cosmological Singularities, }
J. Phys. Conf. Ser. 68 (2007) 012004.


\bibitem{star1}A. A. Starobinsky,  Sov. Phys. JETP Lett. 37 (1983) 66
\bibitem{bcts}J. D. Barrow, S. Cotsakis and A. Tsokaros, Class. Quant. Grav. 27
(2010) 165017.
\bibitem{bctr}J.D. Barrow, S. Cotsakis and D. Trachilis,  Eur. Phys. J. C 80, 1197 (2020); arXiv: 2009.01732
\bibitem{hs}J. Mark Heinzle and P. Sandin, Comm.Math.Phys. 313 (2012) 385; arXiv:1105.1643
\bibitem{ll}L. Landau  and E. M. Lifshitz, \emph{The Classical Theory of Fields}, 4th revised edn (Pergamon, 1975)
\bibitem{kl1}Khalatnikov I M and Lifshitz E M 1961 Sov. Phys. JETP 12 108
\bibitem{kl2} Khalatnikov I M and Lifshitz E M 1964 Sov. Phys. Usp. 6 495
\bibitem{kkms}Khalatnikov I M, A Y Kamenshchik, M Martellini and A A Starobinsky 2003 J. Cosmol. Astropart. Phys. 03 (2003) 001
\bibitem{kks} Khalatnikov I M, A Y Kamenshchik, and A A Starobinsky, Class. Quantum Grav. 19 (2002) 3845
\bibitem{dl}N. Derouelle and D. langlois, Phys. Rev. D52 (1995) 2007
\bibitem{to}K. Tomita, Phys. Rev. D48 (1993) 5634

\bibitem{group1}A. Riess et al, Astron. J. 116 (1998) 1009
\bibitem{group2}S. Perlmutter et al, Ap. J. 517 (1999) 565
\bibitem{copeland}E. J. Copeland et al, \emph{Dynamics of dark energy,} Int.J.Mod.Phys.D 15 (2006) 1753-1936; e-Print: hep-th/0603057 [hep-th]
\bibitem{clifton} T. Clifton et al, \emph{Modified gravity and cosmology,} Phys.Rept. 513 (2012) 1-189; e-Print: 1106.2476 [astro-ph.CO]
\bibitem{capo}S. Capozziello and M.  De Laurentis, \emph{Extended Theories of Gravity,}  Phys.Rept. 509 (2011) 167-321; e-Print: 1108.6266 [gr-qc]
\bibitem{joyce}A. Joyce et al, \emph{Beyond the Cosmological Standard Model,} Phys.Rept. 568 (2015) 1-98; e-Print: 1407.0059 [astro-ph.CO]
    \bibitem{olmo1}J. B. Jimenez et al, \emph{Born–Infeld inspired modifications of gravity,}  Phys.Rept. 727 (2018) 1-129; e-Print: 1704.03351 [gr-qc]
    \bibitem{olmo2}G. J. Olmo, \emph{Palatini Approach to Modified Gravity: $f(R)$ Theories and Beyond,} Int.J.Mod.Phys.D 20 (2011) 413-462
    \bibitem{horn}T. Kobayashi, \emph{Horndeski theory and beyond: a review,} Rept.Prog.Phys. 82 (2019) 8, 086901; e-Print: 1901.07183 [gr-qc]
    \bibitem{eff}N. Frusciante, \emph{Effective field theory of dark energy: A review}, Phys.Rept. 857 (2020) 1-63; e-Print: 1907.03150 [astro-ph.CO]
        \bibitem{holo}S. Wang, Y. Wang, and  M. Li, \emph{Holographic Dark Energy,} Phys.Rept. 696 (2017) 1-57; e-Print: 1612.00345 [astro-ph.CO]
\end{thebibliography}
\end{document}